\documentclass[prd,twocolumn,showpacs,superscriptaddress,floatfix]{revtex4}
\pdfoutput=1

\usepackage{amsfonts}
\usepackage{amsmath}
\usepackage{amssymb}
\usepackage{bm}
\usepackage{dcolumn}
\usepackage{epsfig}
\usepackage[T1]{fontenc}
\usepackage{graphicx}
\usepackage{graphics}
\usepackage{hyperref}
\usepackage[latin1]{inputenc}
\usepackage{latexsym}
\usepackage{multirow}
\usepackage{rotating}

\newcommand\be{\begin{equation}}
\newcommand\ba{\begin{eqnarray}}
\newcommand\ee{\end{equation}}
\newcommand\ea{\end{eqnarray}}

\newcommand{\mb}[1]{\mbox{\boldmath $#1$}}
\newcommand{\salto}[1]{\left[\,#1\,\right]^{}_{p}}
\newcommand{\saltonum}[3]{\left[\,#1\,\right]^{#2,#3}_{p}}
\newcommand{\nn}{\nonumber}
\newcommand{\met}{\mbox{g}}

\newcommand{\rsu}{r^{\ast}}
\newcommand{\rsup}{r^{\ast}_{p}}
\newcommand{\drsup}{\dot{r}^{\ast}_{p}}
\newcommand{\ddrsup}{\ddot{r}^{\ast}_{p}}
\newcommand{\drsupsq}{\dot{r}^{\ast\,2}_{p}}

\newcommand{\singu}{{\mbox{\tiny S}}}
\newcommand{\regu}{{\mbox{\tiny R}}}
\newcommand{\Hor}{{\mbox{\tiny H}}}
\newcommand{\Inf}{{\mbox{\tiny I}}}
\newcommand{\Ai}{{\mbox{\tiny A}}}
\newcommand{\DD}{{\mbox{\tiny D}}}
\newcommand{\LL}{{\mbox{\tiny L}}}
\newcommand{\NN}{{\mbox{\tiny N}}}
\newcommand{\PP}{{\mbox{\tiny P}}}

\newcommand{\MAX}{{\mbox{\tiny max}}}

\begin{document}

%%%%%%%%%%%
%  TITLE  %
%%%%%%%%%%%
\title{Pseudospectral Collocation Methods for the Computation of the Self-Force on a Charged Particle: 
Generic Orbits around a Schwarzschild Black Hole}

%%%%%%%%%%%%%
%  AUTHORS  %
%%%%%%%%%%%%%
\author{Priscilla Canizares}
\affiliation{Institut de Ci\`encies de l'Espai (CSIC-IEEC), 
Facultat de Ci\`encies, Campus UAB, Torre C5 parells, 
E-08193 Bellaterra, Spain}

\author{Carlos F.~Sopuerta}
\affiliation{Institut de Ci\`encies de l'Espai (CSIC-IEEC), 
Facultat de Ci\`encies, Campus UAB, Torre C5 parells, 
E-08193 Bellaterra, Spain}

\author{Jos\'e Luis Jaramillo}
\affiliation{Max Planck Institute f\"ur Gravitationsphysik, 
Albert-Einstein-Institute, 14476 Postdam, Germany}
\affiliation{Laboratoire Univers et Th\'eories (LUTH), Observatoire de Paris, CNRS,
Universit\'e Paris Diderot, 5 place Jules Janssen, 92190 Meudon, France} 

%%%%%%%%%%
%  DATE  %
%%%%%%%%%%
\date{\today}

%%%%%%%%%%%%%%%%%%%%%
%  PREPRINT NUMBER  %
%%%%%%%%%%%%%%%%%%%%%
\preprint{}

%%%%%%%%%%%%%%
%  ABSTRACT  %
%%%%%%%%%%%%%%
\begin{abstract}
The inspiral of a stellar compact object into a massive black hole, an extreme-mass-ratio 
inspiral, is one of the main sources of gravitational waves for the future space-based 
Laser Interferometer Space Antenna.  We expect to be able to 
detect and analyze many cycles of these slowly inspiraling systems, which makes them 
truly high precision tools for gravitational-wave astronomy. To that end, the use of very 
precise theoretical waveform templates in the data analysis is required.  To build them 
we need to have a deep understanding of the gravitational backreaction mechanism responsible 
for the inspiral. The self-force approach describes the inspiral as the action of a local 
force that can be obtained from the regularization of the perturbations created by the 
stellar compact object on the massive black hole geometry.  In this paper we extend a new 
time-domain technique for the computation of the self-force from the circular case to the 
case of eccentric orbits around a non-rotating black hole.  The main idea behind our 
scheme is to use a multidomain framework in which the small compact object, described as 
a particle, is located at the interface between two subdomains.  Then, the equations at 
each subdomain are homogeneous wave-type equations, without distributional sources.  In 
this {\em particle-without-particle} formulation, the solution of the equations is  
smooth enough to provide good convergence properties for the numerical computations.  
This formulation is implemented by using a pseudospectral collocation method for the 
spatial discretization, combined with a Runge Kutta algorithm for the time evolution.  
We present results from several simulations of eccentric orbits in the case of a 
scalar charged particle around a Schwarzschild black hole, an excellent testbed model for 
testing the techniques for self-force computations.  In particular, we show the 
convergence of the method and its ability to resolve the field and its derivatives across 
the particle location. Finally, we provide numerical values of the self-force for 
different orbital parameters.
\end{abstract}

%%%%%%%%%%
%  PACS  %
%%%%%%%%%%

%%%%%%%%%%%%%%%
%  MAKETITLE  %
%%%%%%%%%%%%%%%
\maketitle

%%%%%%%%%%%%%%%%%%
%  INTRODUCTION  %
%%%%%%%%%%%%%%%%%%
\section{Introduction} \label{intro}

One of the main subjects of activity in the emergent area of Gravitational Wave Astronomy
is the physical description of the different astrophysical and cosmological sources of gravitational
radiation that are expected to be detected by the present generation (LIGO~\cite{LIGO}, GEO600~\cite{GEO600}, 
VIRGO~\cite{VIRGO}) and future/planned (LCGT~\cite{LCGT}, LISA~\cite{LISA}, ET~\cite{Einsteint}, etc.) 
laser interferometer observatories.   The description of sources of gravitational waves plays an 
important role for several reasons.  On the one hand, we need to understand well these sources in
order to convert the information from their observations in gravitational waves (and perhaps in
some electromagnetic bands) into new scientific knowledge. On the other hand, in order to detect 
and obtain accurate physical information from these
observations we need to use data analysis techniques, like the \emph{matched filter}, that require the
use of precise waveform templates in order to separate the signals from the noise and to estimate
appropriately the source parameters.

In the case of the future Laser Interferometer Space Antenna (LISA)~\cite{LISA} 
(see~\cite{lobosopuertacqg:2009zz,lobosopuertajpcs:2009zz} for the
proceedings of the 7th international LISA symposium), the main expected
sources of gravitational waves are: Galactic compact binaries, mergers of Massive Black Holes (MBHs),  
the inspirals of stellar-mass compact objects into MBHs, and stochastic gravitational-wave backgrounds
of diverse origin.  This paper is devoted to the description of the dynamics of the third type of
source which, due to the extreme mass ratios involved, are usually known as Extreme-Mass-Ratio 
Inspirals (EMRIs).  The EMRIs of interest for LISA are associated with relaxed galactic nuclei that
have a dense stellar population (a stellar \emph{cusp} or core) around a hypothetical central MBH.  
In this situation, there is a non-negligible probability for a stellar-mass compact object (SCO),
with masses in the range $m\sim 1-50 M^{}_{\odot}$, to end up in a slow inspiral towards the MBH 
(see also~\cite{AmaroSeoane:2007aw} for details and references on the astrophysics of EMRIs).
The range of the MBH mass of interest for LISA is $M \sim 10^4-10^7 M_{\odot}$, which means that the mass
ratios involved in EMRIs are in the interval $\mu = m/M \sim 10^{-7}-10^{-3}$.

In brief, the situation we face in trying to detect EMRIs with LISA is the following: They are systems
that during the last year of inspiral before plunge complete of the order of $10^5$ cycles~\cite{Finn:2000sy},
so they fall into the $\sim 0.1$ mHz part of the LISA band.  Therefore, the mass ratio implies not 
only the existence of very different spatial scales (e.g. the size of the SCO as compared to the size
of the MBH), but also of two very different time scales, namely the orbital time scale versus the
inspiral time scale, so that $T^{}_{\mbox{\tiny orbital}}/T^{}_{\mbox{\tiny inspiral}}\sim \mu = m/M$.  
Despite the difficulties  introduced by the extreme mass
ratios involved in the problem, there is an advantage to it: we can assume that the spacetime geometry
is the one due to the MBH with deviations produced by the SCO.  Then, one can use perturbation theory
to deal with EMRIs.  In this context, the inspiral can be described as the action of a local force,
the so-called {\em self-force}, according to the equation of motion derived in~\cite{Mino:1997nk,Quinn:1997am},
known as the MiSaTaQuWa equation of motion.   The MiSaTaQuWa equation can be used as the foundations of a
self-consistent scheme to simulate an EMRI event in an \emph{adiabatic} way by coupling it to the 
Partial Differential Equations (PDEs) describing the perturbations produced by the SCO (together with
a scheme to identify the regular part of the perturbations responsible for the inspiral).  For recent
studies on these issues see~\cite{Gralla:2008fg,Pound:2009sm,Pound:2010pj}.  For reviews on different
aspects of EMRIs see~\cite{Glampedakis:2005hs,Poisson:2004lr,Barack:2009fk}.

Given the amount of cycles required for EMRIs and their computational cost we cannot expect to generate complete
waveform template banks by means of full self-force calculations.  Instead, the goal of these studies
should be to understand all the details of the structure of the self-force so that we can formulate
efficient and precise algorithms to create the waveforms needed for LISA data analysis, perhaps 
complementing present approximate schemes~\cite{Barack:2003fp,Babak:2006uv,Drasco:2005kz,Yunes:2009ef}.

Moreover, due to the large number of cycles that will be available in future LISA EMRI observations, 
we expect to determine the EMRI physical parameters with high precision~\cite{Barack:2003fp}.
Given also that the expected event rate is in the range $10-10^3$ EMRI$/$yr~\cite{Gair:2004iv,Hopman:2006xn} 
up to a redshift $z\approx 1$, these observation will have an important impact not only on astrophysics
(see~\cite{AmaroSeoane:2007aw} and references therein), but also on cosmology and fundamental 
physics~\cite{Schutz:2009zz}, from precise measurements of cosmological 
parameters~\cite{MacLeod:2007jd,Gair:2008bx} to tests of the geometry of 
MBHs~\cite{Collins:2004ex,Glampedakis:2005cf,Barack:2006pq,Yunes:2007zp,Brink:2008xx,Vigeland:2009pr,Hughes:2010xf} 
and/or the gravitational interaction~\cite{Hughes:2006pm,Sopuerta:2009iy}.

If Intermediate-Mass Black Holes (IMBHs)~\cite{Miller:2003sc,Miller:2008fi} exist 
(with masses in the range $10^2-10^4\,M^{}_{\odot}$), as several recent observations suggest, 
they can form binaries with intermediate mass ratios resulting in the so-called Intermediate-Mass-Ratio 
Inspirals (IMRIs).  LISA will be sensitive to one type of IMRIs: That an IMBH into a MBH.  
In contrast, the inspiral of an SCO into an IMBH, another type of IMRIs outside the LISA band, 
is of interest for the next generations of ground-based detectors~\cite{Brown06,Mandel:2007hi}.
We can obtain an approximate description (probably not accurate enough for data analysis purposes)
of IMRIs by using the same methods that are used for EMRIs.

In the calculation of the self-force, it is reasonable to neglect the structure of the SCO 
and describe it as a point-like object.  This means that the retarded metric pertubations 
generated by the SCO diverge at the particle location, and hence they need to be regularized.
An algorithm that has been widely employed is the {\em mode sum} regularization 
scheme~\cite{Barack:1999wf,Barack:2000eh,Barack:2001bw,Mino:2001mq,Barack:2001gx,Barack:2002mha,Detweiler:2002gi,Haas:2006ne}. 
It substracts, mode by mode (harmonic $\ell$-modes) from the retarded field, the singular part of the perturbations 
that does not contribute to the self-force.  We therefore need a method for computing the different multipoles
of the retarded part of the metric perturbations induced by the SCO.  Due to the complexity of these linear equations,
solutions can only be obtained numerically, either using frequency- or time-domain methods.  Both have advantages and
disadvantages.  Because EMRIs are expected to have high eccentricities and frequency-domain methods
have shown difficulties in this case, we concentrate on time-domain methods.  Time-domain methods have difficulties
related to the resolution of the SCO (see, e.g.~\cite{Sopuerta:2005rd}), since many of them need to introduce an
artificial scale in order to model the distributional energy-momentum tensor that describes the SCO.

To solve the problems related to the artificial scale associated with the SCO some new techniques have been
recently proposed~\cite{Barack:2007jh,Barack:2007we,Vega:2007mc,Lousto:2008mb} which are similar to the puncture method introduced
in numerical relativity~\cite{Brandt:1997tf,Campanelli:2005dd,Baker:2005vv}.  However, they do not completely solve the 
problem of the spatial resolution associated with the source terms.   We recently introduced~\cite{Canizares:2008dp,Canizares:2009ay} 
a new technique, for the case of circular orbits, that completely avoids the introduction of a spatial scale.  The method,
described in detail in~\cite{Canizares:2009ay} (henceforth Paper I), is based on the splitting of the spatial computational
domain (the one-dimensional domain associated with the radial coordinate) into several subdomains, in such a way that the SCO 
is located at the interface of two such subdomains.  For the circular case, the SCO is at rest in the spatial domain.  
The outcome of this setup is that the equations at each subdomain are homogeneous wave-like equations.  In this way,
we ensure the smoothness of the solutions in each domain and at the same time we avoid artificial scales associated with the SCO.
The SCO appears in the boundary/matching conditions between the subdomains that surround it.   This setup is 
ideal to use high-precision numerical methods, which normally require a high degree of smoothness in the solution.
In Paper I we implemented these ideas for the case of a charged scalar particle in circular orbits around a non-rotating
MBH using the PseudoSpectral Collocation (PSC) method for the spatial discretization (a similar technique has also
been proposed recently in~\cite{Field:2009kk}).  In this sense, the smoothness
of the solutions of the homogeneous wave-like equations involved, ensures the exponential convergence of the method, 
as shown in Paper I.  This study has further demonstrated the high precision of the method by comparing results with
previous calculations of the self-force on a charged particle in circular 
motion~\cite{Haas:2007kz,DiazRivera:2004ik,Haas:2006ne}.   Finally, it was shown that by adjusting the subdomains and
the number of collocation points, one also obtains an efficient method (in terms of computational time) for the
calculation of the self-force.  These techniques can be directly imported to the gravitational (and electromagnetic)
case.

The main aim of the present paper is to show that the techniques shown in Paper I can be extended to the case
of generic eccentric orbits.  The key ingredient of our formulation, which we call the Particle-without-Particle
formulation, is to divide the computational domain into subdomains and to locate the particle at the interface of two
of them in such a way that the equations do not have a distributional term that deteriorates the
smoothness of the solution.  The challenge is to ensure communication between the domains in an appropriate way
to take into account the particle effects in a precise way.  The main challenge in comparison with Paper I 
is to adapt these techniques to the case where the particle is no longer at rest in the spatial domain, 
as its radial coordinate is changing with time.  
This will be achieved by adapting the mapping between the physical 
domain (corresponding to the radial coordinate) and the spectral domain (corresponding to the coordinate that labels 
the collocation points used in the PSC method) such that
the particle is moving in the physical domain but it is at rest in the spectral domain.  In this way, we can
apply all the machinery shown in Paper I.  Here we present the ingredients of the new formulation in order 
to handle eccentric orbits: the equations, the junction conditions, 
different ways of implementing it numerically, etc.
In particular we show that the spectral convergence is preserved.  We also show that we can evolve the
system in time with long-term stability and a good resolution of the jumps in the derivatives of the solution induced by
the presence of the particle.  In particular, we provide values of the self-force components, at the pericenter radius, 
for different eccentric orbits.

The organization of this paper is as follows: 
In Sec.~\ref{formulation} we introduce the formulation of the physical model, 
a charged scalar particle orbiting a Schwarzschild MBH, including the basic formulae for the computation 
of the self-force via the mode-sum regularization scheme.
In Sec.~\ref{pwpformulation} we introduce the basic mathematical formulation 
of our method, based on the division of the physical domain into subdomains and on the communication between them.
In Sec.~\ref{pscimplementation} we describe how to implement the mathematical formulation of Sec.~\ref{pwpformulation}
using the PSC method.
In Sec.~\ref{results} we show the performance of the numerical codes that we have designed
to implement the new scheme as well as results from the computation of the self-force.
We conclude with a summary and discussion in Sec.~\ref{discussion}, including future work.
Three appendices complete this paper: In Appendix~\ref{particle_motion} we summarize the main equations
that describe the geodesic motion that we use in this work.  
In Appendix~\ref{singularfield} we provide the
main formulae related to the structure of the singular field and the associated regularization
parameters.  Finally, in Appendix~\ref{jump_derivatives}, we just provide the expressions for the 
time derivatives of the jumps in the characteristic variables.
Throughout this paper we use the metric signature $(-,+,+,+)$ and geometric
units in which $G = c = 1$.

%%%%%%%%%%%%%%%%%%%%%%%%%%%%%%%%%%%%%%%%%%%%%%%%%%%%%%%%%%%%
%   Charged Scalar Particle Orbiting a Schwarzschild MBH   %
%%%%%%%%%%%%%%%%%%%%%%%%%%%%%%%%%%%%%%%%%%%%%%%%%%%%%%%%%%%%
\section{Charged Scalar Particle Orbiting a Schwarzschild MBH}   \label{formulation}

In order to simplify the presentation of our techniques we use a simplified EMRI model,
corresponding to a charged scalar particle orbiting a non-rotating black hole.  In this
model the SCO is represented as a particle, with charge $q$ associated to a 
scalar field $\Phi$, and the MBH is described by a fixed spacetime geometry, in the
sense that it is not affected by the charged particle.  Since we are interested in the 
dynamics of EMRIs in General Relativity, the
geometry of the non-rotating MBH is given by the Schwarzschild spacetime:
\begin{equation}
ds^2=f(-dt^2+d\rsu{}^{2})+r^2d\Omega^2\,,~~
f=1-\frac{2M}{r}\,,
\label{schmetric}
\end{equation}
where $M$ is the MBH mass, $d\Omega^2=d\theta^2+\sin^2\theta\,d\varphi^2$ denotes the
metric of the two-sphere, and
\be
\rsu = r + 2M \ln\left(\frac{r}{2M}-1\right) \,,
\ee
is the so-called radial \emph{tortoise} coordinate.  The particle induces the scalar
field $\Phi$ satisfying the following wave-like equation (see, e.g.~\cite{Poisson:2004lr}):
\begin{equation}
\met^{\alpha\beta}\nabla^{}_{\alpha}\nabla^{}_{\beta}\Phi(x)= -4\pi q\int^{}_{\gamma}
d \tau\, \delta^{}_4(x,\textit{z}(\tau))  \,, \label{geo}
\end{equation}
where $\met^{\alpha\beta}$ denotes the Schwarzschild inverse metric~(\ref{schmetric}) and
$\nabla^{}_{\mu}$ the associated canonical connection; $\tau$ denotes proper time associated 
with the particle along its time-like worldline $\gamma$, with coordinates $x^{\mu} = 
z^{\mu}(\tau)$ and  $\delta^{}_{4}(x,x')$ is the invariant Dirac biscalar distributional functional 
in Schwarzschild spacetime, defined as
\begin{equation}
\left(\delta^{}_{4},f\right)(x) \equiv \int d^{4}x' \sqrt{-\met(x')} f(x')
\delta^{}_{4}(x',x) = f(x)\,,
\end{equation}
with $\met$ being the metric determinant.

The field equation~(\ref{geo}) has to be complemented with the equation of motion for the
scalar charged particle:
\begin{equation}
m\frac{du^{\mu}}{d\tau} = F^{\mu} = q (\met^{\mu\nu} + u^{\mu}u^{\nu}) \left. 
\left(\nabla^{}_{\nu}\Phi\right) \right|^{}_{\gamma} \,, ~~
u^{\mu} = \frac{dz^{\mu}}{d\tau} \,, \label{particlemotion}
\end{equation}
where $m$ and $u^{\mu}$ are the mass and 4-velocity of the particle, respectively. The coupled
set of equations formed by the Partial Differential Equation (PDE) for the scalar field~(\ref{geo})
and the Ordinary Differential Equation (ODE) for the particle trajectory~(\ref{particlemotion})
constitute our testbed model for an EMRI.   The SCO generates scalar field according to (\ref{geo}),
which in turn, affects to the SCO motion according to~(\ref{particlemotion}), that is, through 
the action of a local force.  This mechanism is the analogous of the
gravitational backreaction mechanism that produces the inspiral via the gravitational self-force.  

However, this is not the end of the story since the retarded solution of~(\ref{geo}) is singular
at the particle location, whereas the force in Eq.~(\ref{particlemotion}) involves the gradient
of the field evaluated at the particle location.  Therefore, as they stand, Eqs.~(\ref{geo}) 
and~(\ref{particlemotion}) are formal equations that require an appropriate regularization to 
become fully meaningful.   Following~\cite{Detweiler:2002mi}, the retarded field can be split 
into two parts: a singular piece, $\Phi^{\singu}$, which contains the singular structure of the field 
and satisfies the same wave equation as the retarded field, i.e. Eq.~(\ref{geo}), and a regular part, 
$\Phi^{\regu}$, that satisfies the homogeneous equation associated with Eq.~(\ref{geo}).  
As it turns out, $\Phi^{\regu}$ is regular and differentiable at the 
particle's position and is the solely responsible of the scalar self-force~\cite{Detweiler:2002mi}.  
We can therefore write
\begin{equation}
F_{\regu}^{\mu} =  q (\met^{\mu\nu} + u^{\mu}u^{\nu})\left. 
\left(\nabla^{}_{\nu}\Phi^{\regu}\right) \right|^{}_{\gamma} \,, 
\label{regparticlemotion}
\end{equation}
which gives a definite sense to the equations of motion of the system.

In solving the field equation~(\ref{geo}) we can make use of the spherical symmetry of the
MBH geometry to expand the solution in scalar spherical harmonics (see Paper I for conventions):
\begin{eqnarray}
\Phi = \sum_{\ell=0}^{\infty}\sum_{m=-\ell}^{\ell}\Phi^{\ell m}(t,r)Y^{\ell m}(\theta,\varphi)\;. 
\label{phiex}
\end{eqnarray}
The advantage of this expansion is that each harmonic mode, 
$\Phi^{\ell m}(t,r)$, 
is decoupled from the rest and satisfies the following $1+1$ wave-type equation:
\begin{eqnarray}
\left\lbrace  -\frac{\partial^2}{\partial t^2} + \frac{\partial^2}{\partial \rsu{}^{2}} 
-V^{}_{\ell}(r) \right\rbrace(r\Phi^{\ell m})= S^{\ell m}\delta (r-r^{}_{p}(t))\,, 
\label{master}
\end{eqnarray}
where 
\be
V^{}_{\ell}(r) = f(r) \left[ \frac{\ell(\ell+1)}{r^2}+\frac{2M}{r^3}\right]\,,
\ee
is the Regge-Wheeler potential for scalar fields on the Schwarzschild geometry and
\begin{eqnarray}
S^{\ell m} = -\frac{4\pi qf(r^{}_{p})}{r^{}_{p}u^t}\,\bar{Y}^{\ell m}(\frac{\pi}{2},
\varphi^{}_{p})\,,
\label{source}
\end{eqnarray}
is the coefficient of the singular source term~\footnote[1]{Notice that in Paper I 
there is a typo in the formula for the source term, it contains $f^2(r^{}_{p})$ instead 
of $f(r^{}_{p}).$} due to the presence of the particle,
and the bar denotes complex conjugation.  Here we have assumed, without loss of 
generality, that the particle's orbit takes place in the equatorial plane $\theta=\pi/2$.  
Moreover, $r^{}_{p}$ and $\varphi^{}_{p}$ denote the radial and azimuthal coordinates of 
the particle, and are functions of the coordinate time $t$.

The expansion in spherical harmonics is useful not only for simplifying the construction of
the retarded solution, but also in terms of constructing its regular part, $\Phi^{\regu}$.
Indeed, it turns out that each harmonic mode of the retarded field, $\Phi^{\ell m}(t,r)$, 
is finite and continuous at the location of the particle.  It is the sum over $\ell$ what
diverges there.  Here is where the {\em mode-sum} regularization 
scheme~\cite{Barack:1999wf,Barack:2001gx,Barack:2002mha} comes into play, as it provides
analytic expressions for the singular part of each $\ell$-mode of the retarded field.
These expressions for the singular field are valid only near
the particle location.   The regularized self-force is thus obtained by computing
numerically each harmonic mode of the self-force and subtracting the singular
part provided by the mode-sum scheme.   Thus, the regular part of the gradient of 
the field, $\nabla^{}_{\alpha}\Phi^{\regu}\equiv \Phi^{\regu}_{\alpha}$, is given by
\begin{equation}
\Phi^{\regu}_{\alpha}(\textit{z}^{\mu}(\tau))= \mathop{\lim}\limits_{x^{\mu}\to 
\textit{z}^{\mu}(\tau)}\sum_{\ell=0}^{\infty}\left(  \Phi^\ell_{\alpha}(x^{\mu}) - 
\Phi^{\singu,\ell}_{\alpha}(x^{\mu})\right) \,.
\label{regular2}
\end{equation}
Introducing,
\be
\Phi^\ell_{\alpha}(x^{\mu})=\sum_{m=-\ell}^{\ell}\nabla_{\alpha}(\Phi^{\ell m}(t,r)Y^{\ell m}(\theta,\varphi))\,,
\ee
the structure of the singular field can can be written as:
\begin{eqnarray}
\Phi^{\singu,\ell}_{\alpha} &=& q \left[  \left( \ell + \frac{1}{2} \right) A^{}_{\alpha} 
+ B^{}_{\alpha} + \frac{C^{}_{\alpha}}{\ell+ \frac{1}{2}}\right.  \nonumber \\
&+& \left. \frac{D^{}_{\alpha}}{(\ell- \frac{1}{2})(\ell+ \frac{3}{2})} + \ldots \right] \,.\label{singular}
\end{eqnarray}
The expressions for the regularization parameters, $A^{}_{\alpha}$, $B^{}_{\alpha}$, 
$C^{}_{\alpha}$, and $D^{}_{\alpha}$, can be found in the literature for generic 
orbits~\cite{Barack:2002mha,Kim:2004yi,hoon:2005dh,Haas:2006ne}.  They do not depend on $\ell$ but 
depend on the trajectory of the particle.   The three first coefficients of~(\ref{singular}) are 
responsible for the divergences, whereas the remaining terms converge to zero once they are summed over $\ell$.
The expressions for the regularization parameters that we use in this paper
are listed in Appendix~\ref{singularfield}.

%%%%%%%%%%%%%%%%%%%%%%%%%%%%%%%%%%%%%%%%%%%%%%%%%%%%%%%%%%
% The Particle-without-Particle Formulation              %
%%%%%%%%%%%%%%%%%%%%%%%%%%%%%%%%%%%%%%%%%%%%%%%%%%%%%%%%%%
\section{The Particle-without-Particle Formulation}
\label{pwpformulation}

Before we introduce the multi-domain framework for the numerical calculations that we
present in this paper, it is important first to look at its mathematical foundations.
To that end, we only need to consider two subdomains of the physical domain, 
that is
\be
\rsu \in (-\infty,\infty) = (-\infty,\rsup)\cup (\rsup,+\infty)\,, 
\label{physicalsubdomains}
\ee
where $\rsup = \rsup(t)$ is the radial tortoise coordinate of the particle. 
This particular division obeys to the fact (see Paper I) that the key point of our
technique consists in locating the particle between two subdomains, which 
automatically makes all the wave-type equations governing the dynamics of the
scalar field modes to be homogeneous.  As a direct consequence, we avoid all the 
problems associated with source terms with support at just one
point and prevents us from introducing an artificial scale
associated with the charged particle.   

The problem is thus reduced to the junction conditions for 
the field across the particle.  It is well-known (see, 
e.g.~\cite{Hilbert:1966ch}) that discontinuities in hyperbolic equations (like
the wave equation) can only appear and propagate along characteristics.  In order to
analyze the junction conditions, it is therefore convenient to adopt a hyperbolic formulation
of the equations for the scalar field.  To that end, we introduce the variables
\begin{eqnarray}
\mb{U} = (\psi^{\ell m},\phi^{\ell m},\varphi^{\ell m}) = (r\,\Phi^{\ell m}\,, 
\partial^{}_t\psi^{\ell m}\,, \partial^{}_{\rsu}\psi^{\ell m}) \,.
\label{uvariables}
\end{eqnarray}
at each of the two subdomains in~(\ref{physicalsubdomains}), these satisfy the  
first-order system of PDEs (see~\cite{Sopuerta:2005gz} for an analysis
in second-order form):
\begin{eqnarray}
\partial^{}_t\mb{U}=\mathbb{A}\cdot\partial^{}_{\rsu}\mb{U}+\mathbb{B}\cdot\mb{U}+\mb{S}\,, 
\label{evosystem}
\end{eqnarray}
where the matrices $\mathbb{A}$ and $\mathbb{B}$ are given by 
\begin{eqnarray}
\mathbb{A} = \begin{pmatrix}  0 & 0 & 0 \\ 
                              0 & 0 & 1 \\
                              0 & 1 & 0 \\
\end{pmatrix}\,,~~
\mathbb{B} = \begin{pmatrix}  0 & 1 & 0 \\
                   -V^{}_{\ell} & 0 & 0 \\
                              0 & 0 & 0 \\
\end{pmatrix}\,,
\end{eqnarray}
and the vector $\mb{S}$ is
\be
\mb{S} = \left(0,-\frac{S^{\ell m}}{f(r^{}_{p})}\delta(\rsu-\rsup(t)),0\right)\,.
\label{sourceterm}
\ee
One can check that the system of PDEs given in Eq.~(\ref{evosystem}) constitutes a first-order
symmetric hyperbolic system.  In order to analyze the jumps across the particle in terms
of the variables of Eq.~(\ref{uvariables}), we introduce the following splitting based
on the domain division of Eq.~(\ref{physicalsubdomains}):
\begin{eqnarray}
\mb{U}(t,\rsu) &=& \mb{U}^{}_{-}(t,\rsu)\Theta(\rsup(t)-\rsu) \nonumber \\
&+& \mb{U}^{}_{+}(t,\rsu)\Theta(\rsu - \rsup(t))\,, 
\label{globalsolution}
\end{eqnarray}
where $\mb{U}^{}_{-}$ and $\mb{U}^{}_{+}$ are going to be smooth functions on
$(-\infty,\rsup)$ and on $(\rsup,+\infty)$ respectively, and  $\Theta$ denotes the 
Heaviside step function.  Introducing this ansazt [Eq.~(\ref{globalsolution})] into the 
system~(\ref{evosystem}) and using the following definition of the jump in an 
arbitrary quantity $\lambda$
\be
\salto{\lambda} = \mathop{\lim }\limits_{\rsu \to \rsu_{p}}\lambda^{}_{+}(t, \rsu)-
\mathop{\lim }\limits_{\rsu \to \rsu_{p}}\lambda^{}_{-}(t, \rsu)\,, \label{salto}
\ee
we obtain the evolution equations for $\mb{U}^{}_{\pm}$:
\be
\partial^{}_t\mb{U}^{}_{\pm}=\mathbb{A}\cdot\partial^{}_{\rsu}\mb{U}^{}_{\pm}
+\mathbb{B}\cdot\mb{U}^{}_{\pm}\,, 
\label{homevosystem}
\ee
and the expressions for the jumps in $\mb{U}$ across the particle
(see also~\cite{Sopuerta:2005gz} and Paper I):
\begin{eqnarray}
\salto{\psi^{\ell m}} &=& 0 \,, \label{jump_psi} \\
\salto{\phi^{\ell m}} &=& -\frac{\drsup\,S^{\ell m}}{(1-\drsupsq)f^{}_{p}}\,, \label{jump_phi}\\
\salto{\varphi^{\ell m}} &=& \frac{S^{\ell m}}{(1-\drsupsq)f^{}_{p}}\,, \label{jump_varphi} 
\end{eqnarray}
where the dot denotes differentiation with respect to the time coordinate $t$, $f^{}_{p} =
f(r^{}_{p})$ and $S^{\ell m}$ is given in Eq.~(\ref{source}).  As we can see, the field is 
continuous across the particle, and only the time and radial derivatives of the field have 
a jump, which depend on the particle's speed and satisfy the following advection-like equation:
\begin{eqnarray}
\salto{ \phi^{\ell m} - \drsup\, \varphi^{\ell m} } = 0\,. \label{advection}
\end{eqnarray}
For $\rsup = \mbox{const.}$, i.e. $\drsup=0$,  
we recover the jump conditions for the circular case studied in Paper I.  

As mentioned above, the jump conditions (\ref{jump_psi})-(\ref{jump_varphi}) have to be imposed on 
the characteristic fields because discontinues propagate along the characteristics.
In our case,  the 
characteristic fields are: $\psi^{\ell m}$ (whose characteristic surfaces are the spacelike surfaces 
$\{t=\mbox{const.}\}$), $U^{\ell m} = \phi^{\ell m}-\varphi^{\ell m}$ (whose characteristic surfaces are 
the null surfaces $\{t -\rsu =\mbox{const.}\}$), and $V^{\ell m} = \phi^{\ell m} +\varphi^{\ell m}$ 
(whose characteristic surfaces are the null surfaces $\{t + \rsu =\mbox{const.}\}$). 
See an illustration of this in Figure~\ref{characteristics}.

An alternative description of the system, which we shall also use in this work, is to choose a set 
of variables associated with the characteristic fields.  The obvious choice is
\be
\mb{N} = (\psi^{\ell m},U^{\ell m},V^{\ell m}) \,. \label{charvariables}
\ee
As before, we can apply the same splitting of Eq.~(\ref{globalsolution}) to these variables and follow the
same procedure to obtain a set of equations for $\mb{N}^{}_{\pm}$:
\be
\partial^{}_t\mb{N}^{}_{\pm}=\mathbb{C}\cdot\partial^{}_{\rsu}\mb{N}^{}_{\pm}
+\mathbb{D}\cdot\mb{N}^{}_{\pm}\,, 
\label{homcharevosystem}
\ee
where the matrices $\mathbb{C}$ and $\mathbb{D}$ are
\begin{eqnarray}
\mathbb{C} = \begin{pmatrix}  0 & 0  & 0 \\ 
                              0 & -1 & 0 \\
                              0 & 0  & 1 \\
\end{pmatrix}\,,~~
\mathbb{D} = \begin{pmatrix}  0 & 1/2 & 1/2 \\
                   -V^{}_{\ell} & 0   & 0 \\
                   -V^{}_{\ell} & 0   & 0 \\
\end{pmatrix}\,,
\end{eqnarray}
and the junction condition of Eq.~(\ref{jump_psi}) and
\begin{eqnarray}
\salto{ U^{\ell m} } & = & -\frac{S^{\ell m}}{(1-\drsup)f^{}_{p}}\,,\label{jump_U}  \\  
\salto{ V^{\ell m} } & = & \frac{\drsup\,S^{\ell m}}{(1+\drsup)f^{}_{p}}\,. \label{jump_V}
\end{eqnarray}

We can now establish what we call the \emph{Particle without Particle} (PwP) formulation
of the problem.  By splitting the physical domain from the particle location as in 
Eq.~(\ref{physicalsubdomains}) we can introduce in a natural way the splitting in the
dynamical variables given by Eq.~(\ref{globalsolution})
(and the equivalent one for the characteristic variables $\mb{N}$).  In this way, the 
restriction of the global variables to the subdomains to the left of the particle, 
$(-\infty,\rsup)$, and to the right of the particle, $(\rsup,+\infty)$,  
$\mb{U}^{}_{+}$ ($\mb{N}^{}_{+}$) and $\mb{U}^{}_{-}$ ($\mb{N}^{}_{-}$)
respectively, satisfy
homogeneous hyperbolic equations~(\ref{homevosystem}) [Eq.~(\ref{homcharevosystem}) for 
the characteristic variables].  This means that we have got rid of the distributional 
source terms, due to the particle, that appeared in the global equations~(\ref{evosystem}).  
Then, the presence of the particle is introduced in the communication
between the variables of different subdomains through the junction conditions 
(\ref{jump_psi})-(\ref{jump_varphi}) [Eqs.~(\ref{jump_psi}), (\ref{jump_U}), and (\ref{jump_V})
for the characteristic variables].

%%%%%%%%%%%%%%%%%%%%%%%%%%%%%%%%%%%%%%%%%%%%%%%%%%%%%%%%%%
% A PSC Method Implementation of the PwP Scheme          %
%%%%%%%%%%%%%%%%%%%%%%%%%%%%%%%%%%%%%%%%%%%%%%%%%%%%%%%%%%
\section{A PSC Method Implementation of the PwP Scheme} \label{pscimplementation}
We now introduce a multidomain computational method that implements the PwP scheme described
in the previous section.  To that end, we follow the discussion of Paper I, generalizing
the methods introduced there to include the case of generic orbits, in which the particle is not at 
rest in the physical domain. 

\begin{figure}[htp]
\centering
\includegraphics[width=0.49\textwidth]{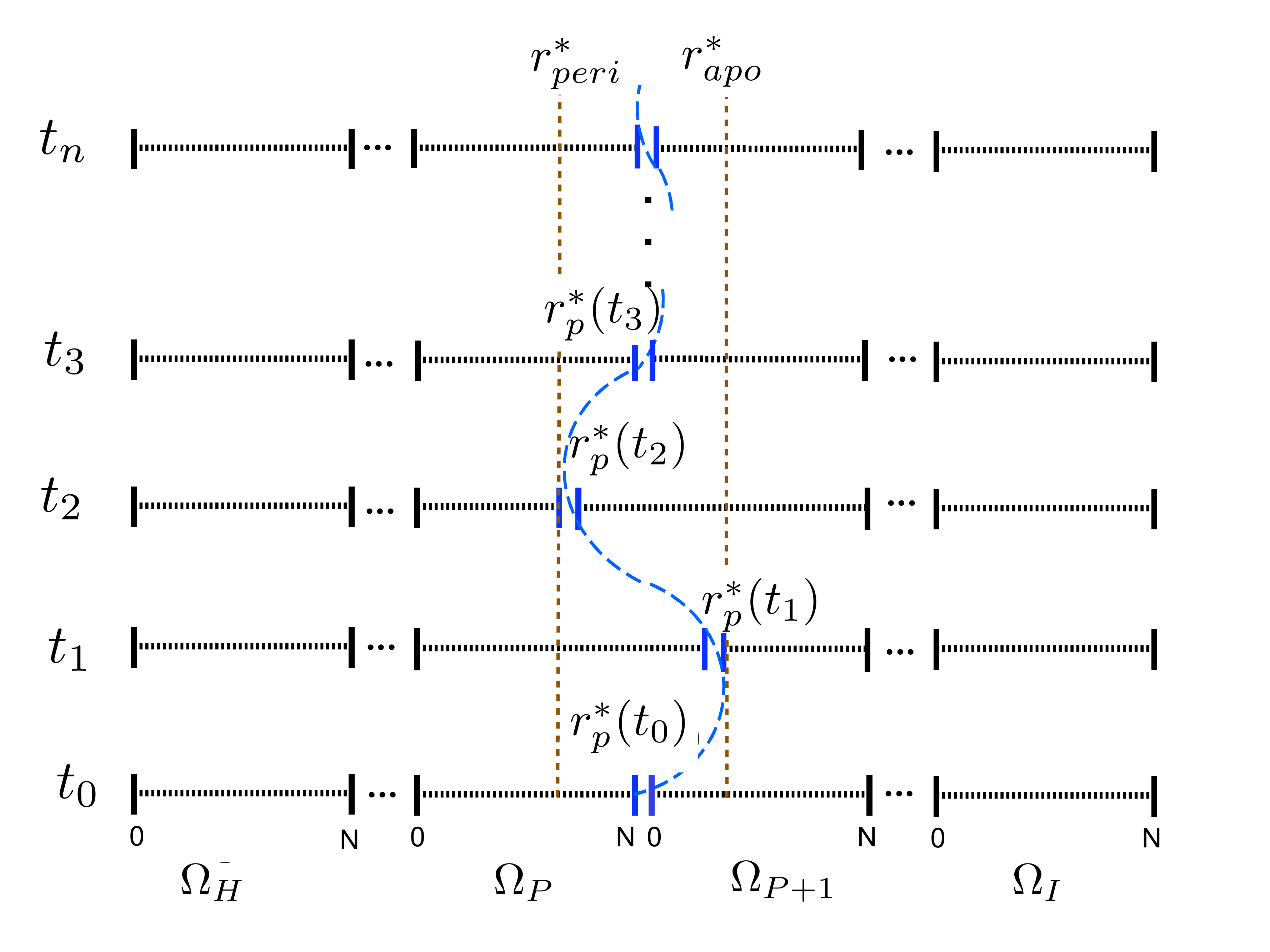}
\caption{The figure shows the structure of the one-dimensional spatial grid, the division in 
subdomains and the location of the particle at the interface between two of them, $\rsup(t)$. 
In a generic orbit, the particle moves in a periodic way between the minimum (pericenter,
denoted by $r^{}_{peri}$) and the maximum (apocenter, denoted by $r^{}_{apo}$).  In this
example, only the domains on the left and on the right of the particle are dynamical, i.e.
they are the only ones whose coordinate size changes in time.}\label{multidomain}
\end{figure}

The first step is the truncation of the physical domain (as opposed to compactification).
Instead of considering the domain~(\ref{physicalsubdomains}) we consider the following
domain: $\Omega=\left[\rsu_{\Hor},\rsu_{\Inf}\right]$, where $\rsu_{\Hor}$ corresponds
to the truncation in the direction to the MBH horizon (at $\rsu\rightarrow -\infty$) and
$\rsu_{\Inf}$ corresponds to the truncation in the direction to spatial infinity 
($\rsu\rightarrow +\infty$).  Next, we divide this truncated 
domain into a number $D$ of subdomains:
\begin{equation}
\Omega = \bigcup^{D}_{\Ai=1} \Omega^{}_{\Ai}\,, ~~\Omega^{}_{\Ai} 
= \left[ \rsu_{\Ai,\LL}, \rsu_{\Ai,\regu}\right]\,,
\end{equation}
where $\rsu_{\Ai,\LL}$ and $\rsu_{\Ai,\regu}$ are the left and right boundaries of the subdomain 
$\Omega^{}_{\Ai}$.   Following the PwP scheme we place the particle at the 
interface between two of the subdomains.   Now, let us assume that the particle is located at 
the interface between the subdomains $\Omega^{}_{\PP}$ and $\Omega^{}_{\PP+1}$, then
\be
\rsu_{p} = \rsu_{\PP,\regu} = \rsu_{\PP+1,\LL}\,. \label{particleatinterface}
\ee
The main difference with respect to the circular case of Paper I is that now $\rsup$ is 
a function of the time $t$.  This implies that $\rsu_{\PP,\regu}$ and $\rsu_{\PP+1,\LL}$
must be functions of $t$ as well, and  the 
subdomains will in general change with time.  We do not need that all subdomains
change with time however, so we distinguish between dynamical and non-dynamical subdomains
according to whether or not any of their boundary nodes moves with time.   This multidomain 
framework is illustrated in Fig.~\ref{multidomain}.   In the remainder of this work we restrict
ourselves to the case in which only the two subdomains that surround the particle are dynamical,
i.e. only $\rsu_{\PP,\regu}$ and $\rsu_{\PP+1,\LL}$ are time dependent.

%%%%% Figure 1: Multidomain structure
%%%%%

The first ingredient in applying the PSC method is the discretization of the spatial domain
(see~\cite{Pfeiffer:2002wt} for a multidomain PSC method for elliptic equations).
In our scheme, each subdomain is discretized independently, by using a 
{\em Lobatto-Chebyshev} grid (see Paper I for details
on our particular implementation, and e.g.~\cite{Boyd} for general details on the method), 
with collocation points given by
\be  
X^{}_{i} = - \cos\left(\frac{\pi\,i}{N}\right) \qquad (i=0,\ldots,N)\,.
\label{lobattochebyshev}
\ee
In order to complete the discretization of the physical domain, we need to specify how the
\emph{spectral} coordinate $X$, associated with the Chebyshev polynomials, is related 
to the \emph{physical} coordinate $\rsu$. 
This is the key ingredient in our scheme; it is here where we establish how to keep the 
particle at the interface of two subdomains even when it is moving in the radial direction.  
This information is encoded in the mapping from the spectral and physical domains.  
We have effectively traded particle motion for coordinate mappings.
Given that the particle will be in general moving, it is convenient to prescribe a 
spacetime mapping.  Since we discretize each such subdomain independently, and 
the spectral domains are always the interval $[-1,1]$, there are going to be as many mappings as subdomains.  
Then, for a given  subdomain $\Omega_{\Ai} =\left[\rsu_{\Ai,\LL}, \rsu_{\Ai,\regu}\right]$ 
($A=0,\ldots,D$), the spacetime mapping to the spectral domain is taken to be 
linear and given by:
\begin{eqnarray}
\begin{array}{ccccc}
X^{}_{\Ai} & : & \mathbb{T}\times\left[ \rsu_{\Ai,\LL}, \rsu_{\Ai,\regu}\right] & \longrightarrow & \mathbb{T}\times\left[-1,1\right] \\[2mm]
           &   & (t,\rsu)                                 & \longmapsto     &  (T, X^{}_{\Ai}) 
\end{array} 
\end{eqnarray} 
where $\mathbb{T} = [t^{}_{o},t^{}_{f}]$ is the evolution time interval and
\begin{eqnarray}
T(t) &=&t\;,\nonumber\\
X^{}_{\Ai}(t,\rsu) &=& \frac{2\rsu- \rsu_{\Ai,\LL}- \rsu_{\Ai,\regu}}{ \rsu_{\Ai,\regu}-\rsu_{\Ai,\LL} } \,.    
\label{map1}
\end{eqnarray} 
Here, the time dependence in $X^{}_{\Ai}$ comes from the variation of $\rsu_{\Ai,\LL}$
or $\rsu_{\Ai,\regu}$ in time. The inverse (linear) mappings from the spectral domain
to each of the subdomains $\Omega_{\Ai}$ are given by
\begin{eqnarray}
\begin{array}{ccccc}
\left.\rsu\right|^{}_{\Omega^{}_{\Ai}} & : & \mathbb{T}\times\left[-1,1\right] & \longrightarrow & \mathbb{T}\times
\left[\rsu_{\Ai,\LL},\rsu_{\Ai,\regu}\right] \\[2mm]
                                     &   & (T, X)                            & \longmapsto     & (t,\rsu) 
\end{array} 
\end{eqnarray}
and
\begin{eqnarray}
t(T) & = & T \,, \nonumber\\
\left.\rsu(T, X)\right|^{}_{\Omega^{}_{\Ai}} & = & 
\frac{\rsu_{\Ai,\regu}-\rsu_{\Ai,\LL}}{2}X+\frac{\rsu_{\Ai,\LL}+ \rsu_{\Ai,\regu}}{2}\,.\label{map2}
\end{eqnarray}

In the PSC method we have two expansions (representations) of our variables: The 
\emph{spectral} and the \emph{physical} representations.  The former expands the variables 
in a basis of Chebyshev 
polynomials, $\{T^{}_{n}(X)\}$ ($X \in [-1,1]$, $n=0,\ldots,N$). Thus, our variables
$\mb{U}$ can be approximated by
\begin{equation}
\mb{U}^{}_{N}(t,\rsu) = \sum_{n=0}^N \mb{a}^{}_n(t)\, T^{}_n(X^{}_\Ai(t,\rsu)) \,,
\label{spectralrepresentation}
\end{equation}
where the $\mb{a}^{}_n$ are the spectral coefficients which, in our case, will be 
time dependent.  The physical representation 
approximates the solution as 
\begin{equation}
\mb{U}^{}_{N}(t,\rsu) = \sum_{i=0}^N \mb{U}^{}_i(t)\, {\cal C}^{}_i(X^{}_\Ai(t,\rsu))\,,
\label{physicalrepresentation}
\end{equation}
where ${\cal C}^{}_i(X)$ are the {\em cardinal} functions associated with our choice
of collocation grid (Lobatto-Chebyshev) which satisfy the relation 
${\cal C}^{}_i(X^{}_{j}) = \delta^{}_{ij}$.  In this representation, the coefficients 
$\{\mb{U}^{}_{i}\}^{}_{i=0,\ldots,N}$ are the values of our variables
\begin{equation}
\mb{U}(t,\rsu(T(t), X^{}_{i})) \equiv \mb{U}^{}_{i}(t) \qquad (i=0,\ldots,N)\,,
\end{equation}
at the collocation points.

In these two representations we have two types of unknowns, the coefficients
$\{\mb{a}^{}_n\}^{}_{n=0,\ldots,N}$ and the collocation values $\{\mb{U}^{}_{i}\}^{}_{i=0,\ldots,N}$.
These $N+1$ unknowns are fully determined, at each subdomain, by imposing our set of PDEs~(\ref{homevosystem})
[or Eq.~(\ref{homcharevosystem})] at each of the collocation points.

Then, the discretized version of the evolution equations (\ref{homevosystem}) at each subdomain  $\Omega_{\Ai}$ 
takes the following form: 
\be
\partial_{_T}\mb{U}^{}_{i} = \mathbb{P}^{}_{i} \,\left(\partial^{}_{X}\mb{U}\right)^{}_{i}
+\mathbb{Q}^{}_{i}\, \mb{U}^{}_{i}\,, \label{phys_equ2}
\ee
where there is no summation over $i$ and the matrices $\mathbb{P}^{}_{i}$ and $\mathbb{Q}^{}_{i}$ are given by
\begin{eqnarray}
\mathbb{P}^{}_{i} = \begin{pmatrix}
0 & 0  & 0 \\[1mm]
0 & -\left(\partial^{}_t X^{}_{\Ai}\right)^{}_{i} & \left(\partial^{}_{\rsu}X^{}_{\Ai}\right)^{}_{i} \\[1mm]
0 & \left(\partial^{}_{\rsu} X^{}_{\Ai}\right)^{}_{i} & -\left(\partial^{}_t X^{}_{\Ai}\right)^{}_{i}
\end{pmatrix}\,, \label{P_matrix}
\end{eqnarray}
\begin{eqnarray}
\mathbb{Q}^{}_{i} = \begin{pmatrix}
0 & 1 & -\frac{\left(\partial^{}_t X^{}_{\Ai}\right)^{}_{i}}{\left(\partial^{}_{\rsu}X^{}_{\Ai}\right)^{}_{i}}\\[2mm]
-V^{\Ai,i}_{\ell} & 0 & 0 \\
0 & 0 & 0 \end{pmatrix}\,, \label{Q_matrix}
\end{eqnarray}
the quantity $V^{\Ai,i}_{\ell}$ is given by
\be
V^{\Ai,i}_{\ell}(t) = V^{}_{\ell}(\rsu(T(t),X^{}_{\Ai,i}))\,,
\ee
$X^{}_{\Ai,i}$ is the $i$-th collocation point at the domain $\Omega_{\Ai}$, given by Eq.~(\ref{lobattochebyshev}),
and 
\be
\left(\partial^{}_{\rsu} X^{}_{\Ai}\right)^{}_{i} = \frac{2}{\rsu_{\Ai,\regu} - \rsu_{\Ai,\LL}}\,.
\label{dX_dr}
\ee
\be
\left(\partial^{}_t X^{}_{\Ai}\right)^{}_{i} = 
-\frac{\drsup}{\rsu_{\Ai,\regu} - \rsu_{\Ai,\LL}}\left\{ \delta^{\Ai,\LL}_{p} + \delta^{\Ai,\regu}_{p} 
+ \left(\delta^{\Ai,\regu}_{p} - \delta^{\Ai,\LL}_{p}\right) X^{}_{\Ai,i} \right\}\,, \label{dX_dt}
\ee
\be
\delta^{\Ai,\LL}_{p} = \left\{ \begin{array}{ll} 1 & \mbox{if~} \rsup = \rsu_{\Ai,\LL}\,, \\
                                               0 & \mbox{otherwise,}
                             \end{array} \right.~
\delta^{\Ai,\regu}_{p} = \left\{ \begin{array}{ll} 1 & \mbox{if~} \rsup = \rsu_{\Ai,\regu}\,, \\
                                               0 & \mbox{otherwise.}
                             \end{array} \right.
\ee

For a given set of initial data
\be 
\mb{U}^{o}_{i} = \mb{U}(t=t^{}_{o},\rsu(T,X^{}_{\Ai,i}))\,,
\ee
the system of $N+1$ ODEs defined by Eq.~(\ref{phys_equ2}) determines uniquely
the values of our variables $\mb{U}$ at the collocation points, i.e.~the $N+1$ unknowns, which in turn 
determines 
uniquely our approximation to the solution via Eq.~(\ref{physicalrepresentation}).    
Looking at the expressions for the matrices $\mathbb{P}^{}_{i}$ and $\mathbb{Q}^{}_{i}$
[Eqs.~(\ref{P_matrix}) and~(\ref{Q_matrix})] we see that their components are explicitly time dependent
in the case of a dynamical subdomain (in the case of this paper, only those that surround the particle).

A similar system of equations can be obtained for the characteristic variables $\mb{N}$ 
[Eq.~(\ref{charvariables})],
\begin{eqnarray}
\partial_{_T}\mb{N}^{}_{i} = \mathbb{R}^{}_{i} \,\left(\partial^{}_{X}\mb{N}\right)^{}_{i}
+\mathbb{S}^{}_{i}\;\mb{N}^{}_{i}\,,\label{phys_equ_char}
\end{eqnarray}
where again there is no summation over $i$ and the matrices $\mathbb{R}^{}_{i}$ and $\mathbb{S}^{}_{i}$ are
\begin{eqnarray}
\mathbb{R}^{}_{i} = \begin{pmatrix}
0 & 0  & 0 \\[1mm]
0 & -\left(\partial^{}_t X^{}_{\Ai}\right)^{}_{i} - \left(\partial^{}_{\rsu}X^{}_{\Ai}\right)^{}_{i} & 0 \\[1mm]
0 & 0 & -\left(\partial^{}_t X^{}_{\Ai}\right)^{}_{i} + \left(\partial^{}_{\rsu} X^{}_{\Ai}\right)^{}_{i} 
\end{pmatrix}\,, \label{R_matrix}
\end{eqnarray}
\begin{eqnarray}
\mathbb{S}^{}_{i} = \begin{pmatrix}
0 & \frac{1}{2}\left[1 +\frac{\left(\partial^{}_t X^{}_{\Ai}\right)^{}_{i}}{\left(\partial^{}_{\rsu}X^{}_{\Ai}\right)^{}_{i}}\right] & 
\frac{1}{2}\left[1 -\frac{\left(\partial^{}_t X^{}_{\Ai}\right)^{}_{i}}{\left(\partial^{}_{\rsu}X^{}_{\Ai}\right)^{}_{i}}\right]\\[2mm]
-V^{\Ai,i}_{\ell} & 0 & 0 \\
-V^{\Ai,i}_{\ell} & 0 & 0 \end{pmatrix}\,, \label{S_matrix}
\end{eqnarray}

However, the ODEs given in Eq.~(\ref{phys_equ2}) do not constitute a complete description
of our system.  We still need to incorporate all the necessary
boundary conditions to adequately describe the system.  First of all, we have the \emph{global}
boundary conditions, at $\rsu_{\Hor}$ and $\rsu_{\Inf}$ to prevent incoming
signals from outside the physical domain, $\Omega=\left[\rsu_{\Hor},\rsu_{\Inf}\right]$, sometimes
known as absorbing boundary conditions.  In this work we try to push the boundaries as far as possible
(typically in such a way that, given the evolution time, $t^{}_{f}-t^{}_{o}$, the boundaries are not in 
causal contact with the particle) and impose Sommerfeld outgoing boundary conditions
\begin{eqnarray}
\phi^{\ell m}(t,\rsu_{\Hor}) - \varphi^{\ell m}(t,\rsu_{\Hor}) & = & 0\,, \label{outgoingbcs1} \\[2mm]
\phi^{\ell m}(t,\rsu_{\Inf}) + \varphi^{\ell m}(t,\rsu_{\Inf}) & = & 0\,. \label{outgoingbcs2}
\end{eqnarray}
These are exact boundary conditions only when $V^{}_{\ell}=0$ or $\rsu_{\Hor}\rightarrow - \infty$ 
and $\rsu_{\Inf}\rightarrow + \infty$.

The second type of boundary conditions that need to be imposed are the junction conditions between the
different subdomains.  
In this paper we present two different methods of imposing them: (i) The \emph{penalty}
method, which has already been used in Paper I, and (ii) direct communication of the characteristic fields.
In the remainder of this section we describe both methods and the modifications that have to be included
for their implementation in the system
of ODEs~(\ref{phys_equ2}).   We recall that in this paper we consider the case in which the only dynamical subdomains
are those two surrounding the particle, $\Omega^{}_{\PP}$ and $\Omega^{}_{\PP+1}$.  
In this way the particle is located at their interface as described by Eq.~(\ref{particleatinterface}). 
We also have:  $\rsu_{\Ai,\regu}=\rsu_{\Ai,\NN}$ and $\rsu_{\Ai+1,\LL} = \rsu_{\Ai+1,0}$.

%%%%%%%%%%%%%%%%%%%%%%%%%%
%   The Penalty method   %
%%%%%%%%%%%%%%%%%%%%%%%%%%
\subsection{The Penalty method} \label{penalty}

The basic idea behind the penalty method is to modify the equations by adding terms that dynamically
drive the system to a state in which the junction conditions are satisfied to a high degree of 
precision. These additional terms are proportional to the conditions that one is interesting in imposing.
In our case they are proportional to the junction conditions.   Details of the method can be found
in Paper I (for generalities see, e.g.~\cite{Hesthaven:2000jh}).  

\begin{figure}[htp]
\centering
\includegraphics[width=0.45\textwidth]{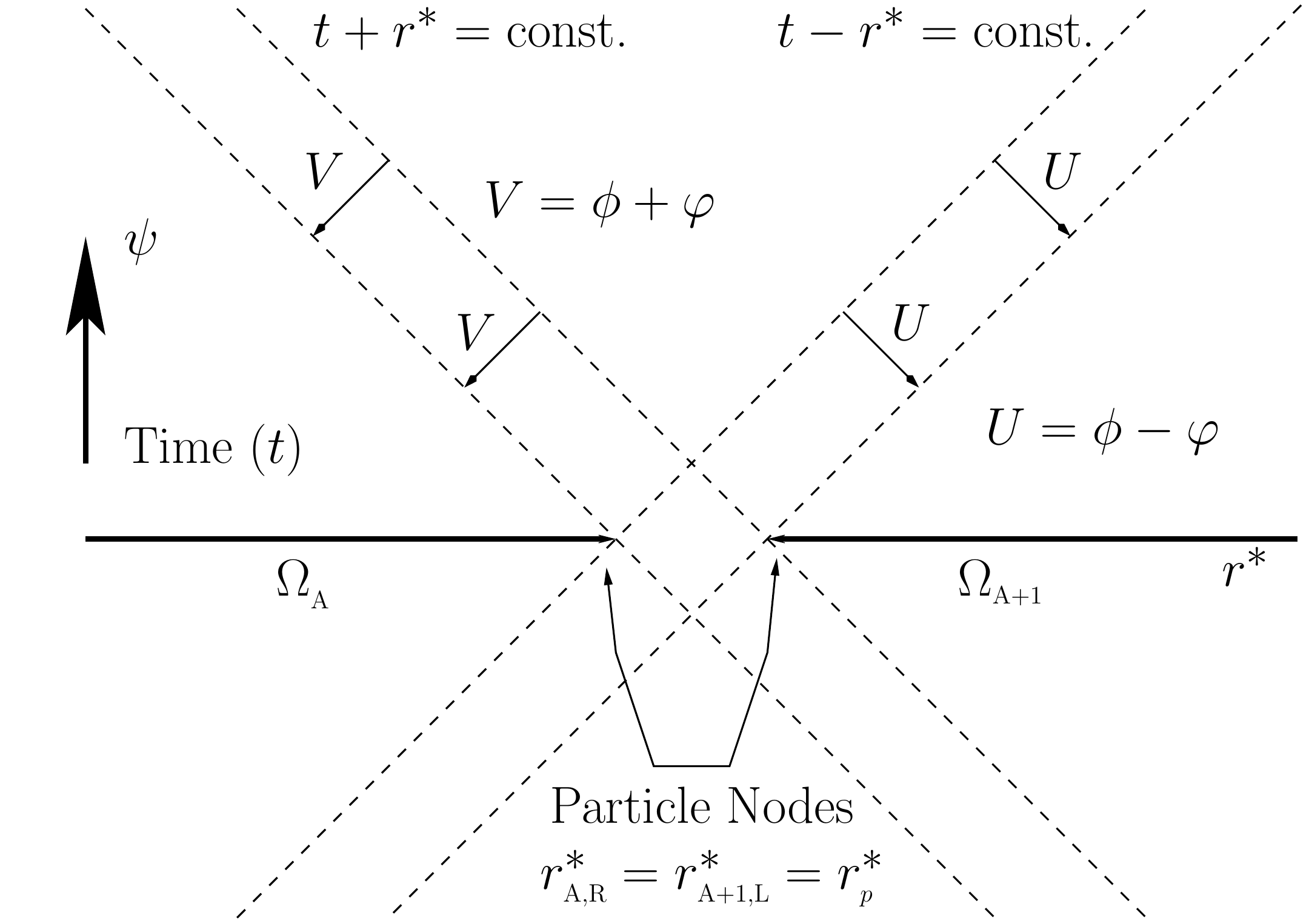}
\caption{Characteristic structure of our field equations. The picture shows the characteristics
($t \pm \rsu = \mbox{const.}$) of
the hyperbolic equations that describe the system [Eq.~(\ref{evosystem})], the characteristic fields ($\psi^{\ell m}$, 
$U^{\ell m}$ $=$ $\phi^{\ell m}$ $-$ $\varphi^{\ell m}$, and $V^{\ell m}$ $=$ $\phi^{\ell m}$ $+$ 
$\varphi^{\ell m}$), and their propagation direction.}\label{characteristics}
\end{figure}

An important point in using the penalty method is that since we are dealing with hyperbolic equations,
the junction conditions have to be imposed across the characteristics (see Figure~\ref{characteristics}), 
and this determines which combinations of the junction conditions have to be used in each equation.

%%%%% Figure 2: Characteristics
%%%%%

In what follows we list the different equations for the different cases that appear depending on whether
we are dealing with a subdomain to the left or to the right of the particle, and also on whether the variables are 
associated with boundary points or not.  That is, we have four basic cases (other possibilities can be derived
from these cases): (i) Inner points for the subdomain to the left of the particle, i.e. $\Omega_{\PP}$; 
(ii) inner points for a subdomain to the right of the particle, i.e. $\Omega_{\PP+1}$; 
(iii) left boundary point; and (iv) right boundary point.

(i) The equations for inner points ($i=1,\ldots,N-1$) valid at the dynamical subdomain at the left of the particle (to simplify the
notation we have dropped the harmonic indices $\ell$ and $m$) are:
\begin{eqnarray}
\partial_{_T}\psi^{}_{\PP,i} & = & \drsup\frac{1 + X^{}_{\PP,i}}{2}\; \varphi^{}_{\PP,i} + \phi^{}_{\PP,i} \,, \label{ileft1} \\
\partial_{_T}\phi^{}_{\PP,i} & = & \drsup\frac{1 + X^{}_{\PP,i}}{2}\left(\partial^{}_{\rsu} \phi^{}_{\PP}\right)^{}_{i}
                                                   + \left(\partial^{}_{\rsu}\varphi^{}_{\PP}\right)^{}_{i} \nn \\
                            & - & V^{\PP,i}_{\ell} \,\psi^{}_{\PP,i}\,, \label{ileft2} \\
\partial_{_T}\varphi^{}_{\PP,i} & = & \drsup\frac{1 + X^{}_{\PP,i}}{2} \left(\partial^{}_{\rsu}\varphi^{}_{\PP}\right)^{}_{i}
                                                       +\left(\partial^{}_{\rsu}\phi^{}_{\PP}\right)^{}_{i} \,, \label{ileft3}
\end{eqnarray}
where for any quantity ${\cal A}$ we compute the spatial derivatives as
\be
\left(\partial^{}_{\rsu} {\cal A}^{}_{\PP}\right)^{}_{i} = 
\left(\partial^{}_{\rsu} X^{}_{\PP}\right)^{}_{i}\left(\partial^{}_{X} {\cal A}^{}_{\PP}\right)^{}_{i} \,,
\ee
and $\left(\partial^{}_{\rsu} X^{}_{\PP}\right)^{}_{i}$ is given in Eq.~(\ref{dX_dr}).

(ii) Equations for inner points at the dynamical subdomain at the right of the particle:
\begin{eqnarray}
\partial_{_T}\psi^{}_{\PP+1,i} & = & \drsup\frac{1 - X_{\PP+1,i}}{2}\; \varphi^{}_{\PP+1,i} + \phi^{}_{\PP+1,i} \,,  \label{iright1} \\
\partial_{_T}\phi^{}_{\PP+1,i} & = & \drsup\frac{1 - X_{\PP+1,i}}{2}\left(\partial^{}_{\rsu}\phi^{}_{\PP+1}\right)^{}_{i}
                                                   + \left(\partial^{}_{\rsu}\varphi^{}_{\PP+1}\right)^{}_{i} \nn \\
                               & - & V^{\PP+1,i}_{\ell} \,\psi^{}_{\PP+1,i}\,, \label{iright2} \\ 
\partial_{_T}\varphi^{}_{\PP+1,i} & = & \drsup\frac{1 - X_{\PP+1,i}}{2}\left(\partial^{}_{\rsu}\varphi^{}_{\PP+1}\right)^{}_{i}
                                                       +\left(\partial^{}_{\rsu}\phi^{}_{\PP+1}\right)^{}_{i} \,. \label{iright3}
\end{eqnarray}
The equations at the inner points of any non-dynamical subdomain are obtained by setting $\drsup = 0$ 
either at the set of equations~(\ref{ileft1})-(\ref{ileft3}) or at the set of equations~(\ref{iright1})-(\ref{iright3}).

(iii) Equations valid at the right boundary of the dynamical subdomain at the left of the particle, that is, 
at $\rsu = r^{\ast}_{\PP,\NN} = \rsup(t)$:
\begin{eqnarray}
\partial_{_T}\psi^{}_{\PP,\NN} & = & \drsup\,\varphi^{}_{\PP,\NN} +  \phi^{}_{\PP,\NN} \nonumber \\
                               & - & \tau^{\PP,\regu}_{\psi}\left( \psi^{}_{\PP,\NN} - \psi^{}_{\PP+1,0}\right)\,, \label{L_psi}\\
\partial_{_T}\phi^{}_{\PP,\NN} & = & \drsup\left(\partial^{}_{\rsu}\phi^{}_{\PP}\right)_{\NN}
                                         + \left(\partial^{}_{\rsu}\varphi^{}_{\PP}\right)_{\NN} 
                                         - V^{\PP,\NN}_{\ell} \,\psi^{}_{\PP,\NN}  \nonumber \\
                                   & - & \frac{\tau^{\PP,\regu}_{\phi}}{2}\left[ \phi^{}_{\PP,\NN} 
                                   + \varphi^{}_{\PP,\NN} - \left(\phi^{}_{\PP+1,0}+ \varphi^{}_{\PP+1,0}\right) \right. \nn \\
                                   & + &      \left. \saltonum{\phi}{\PP}{\PP+1} + \saltonum{\varphi}{\PP}{\PP+1} \right]\,, \label{L_phi}\\
\partial_{_T}\varphi^{}_{\PP,\NN} & = & \drsup\left(\partial^{}_{\rsu}\varphi^{}_{\PP}\right)_{\NN}
                                            + \left(\partial^{}_{\rsu}\phi^{}_{\PP}\right)_{\NN} \nonumber \\ 
                                      & - &  \frac{\tau^{\PP,\regu}_{\varphi}}{2}\left[ \phi^{}_{\PP,\NN} + \varphi^{}_{\PP,\NN} - \left(\phi^{}_{\PP+1,0} 
                                              + \varphi^{}_{\PP+1,0} \right) \right. \nn \\
                                      & + & \left.  \saltonum{\phi}{\PP}{\PP+1} + \saltonum{\varphi}{\PP}{\PP+1} \right]\,, \label{L_varphi}
\end{eqnarray}
where $\tau^{\PP,\regu}_{\psi}$, $\tau^{\PP,\regu}_{\phi}$, and $\tau^{\PP,\regu}_{\varphi}$, are penalty (constant) coefficients
associated with the evolution of $\psi^{}_{\PP,\NN}(T)$, $\phi^{}_{\PP,\NN}(T)$, and $\varphi^{}_{\PP,\NN}(T)$ respectively.
The quantities $\saltonum{\phi}{\PP}{\PP+1}$ and $\saltonum{\varphi}{\PP}{\PP+1}$ are the jumps across the subdomains
$\Omega_{\PP}$ and $\Omega_{\PP+1}$, given in Eqs.~(\ref{jump_phi}) and~(\ref{jump_varphi}).

(iv) Equations valid at the left boundary of the dynamical subdomain at the right of the particle, i.e. at
$\rsu = r^{\ast}_{\PP+1,0}=\rsup(t)$:
\begin{eqnarray}
\partial_{_T}\psi^{}_{\PP+1,0} & = & \drsup\,\varphi^{}_{\PP+1,0} + \phi^{}_{\PP+1,0} \nn \\
                               & - & \tau^{\PP+1,\LL}_{\psi} \left( \psi^{}_{\PP+1,0} - \psi^{}_{\PP,\NN} \right)\,,\label{R_pshi} \\
\partial_{_T}\phi^{}_{\PP+1,0} & = & \drsup\left(\partial^{}_{\rsu}\phi^{}_{\PP+1}\right)^{}_{0} 
                                          +\left(\partial^{}_{\rsu}\varphi^{}_{\PP+1}\right)^{}_{0} 
                                          - V^{\PP+1,0}_{\ell}\,\psi^{}_{\PP+1,0} \nonumber \\
                               & - & \frac{\tau^{\PP+1,\LL}_{\phi}}{2}\left[ \phi^{}_{\PP+1,0} - \varphi^{}_{\PP+1,0} - 
                                     \left( \phi^{}_{\PP,\NN} - \varphi^{}_{\PP,\NN}\right) \right. \nonumber \\
                               & - &  \left. \saltonum{\phi}{\PP}{\PP+1} + \saltonum{\varphi}{\PP}{\PP+1} \right] \,, \label{R_phi}\\
\partial_{_T}\varphi^{}_{\PP+1,0} & = & \drsup\left(\partial^{}_{\rsu}\varphi_{\PP+1}\right)^{}_{0} 
                                             +\left(\partial^{}_{\rsu}\phi^{}_{\PP+1}\right)^{}_{0} \\ 
                                  & + & \frac{\tau^{\PP+1,\LL}_{\varphi}}{2}\left[ \phi^{}_{\PP+1,0} - \varphi^{}_{\PP+1,0}
                                        -\left(\phi^{}_{\PP,\NN}-\varphi^{}_{\PP,\NN}\right) \right. \nonumber \\
                                & - & \left. \saltonum{\phi}{\PP}{\PP+1} + \saltonum{\varphi}{\PP}{\PP+1} \right] \,,  \label{R_varphi}
\end{eqnarray}
where again $\tau^{\PP+1,\LL}_{\psi}$, $\tau^{\PP+1,\LL}_{\phi}$, and $\tau^{\PP+1,\LL}_{\varphi}$ are penalty 
coefficients associated with the evolution of $\psi^{}_{\PP+1,0}(T)$, $\phi^{}_{\PP+1,0}(T)$, and $\varphi^{}_{\PP+1,0}(T)$ respectively.
At this point we comment on the equation for $\psi^{\ell m}$.  Here, we are enforcing the continuity of this variable via a 
penalty term.  However, as we already mentioned in Paper I, there is an alternative to this, which consists in taking a
sort of average of the right-hand side of the equation for $\psi^{\ell m}$ over the two subdomains that we are trying to
communicate or match.  The alternative equation for $\psi^{\ell m}$ would look as follows:
\be
\partial_{_T}\psi^{}_{\PP+1,0} = \frac{1}{2}\drsup\,\left(\varphi^{}_{\PP,\NN}+\varphi^{}_{\PP+1,0}\right) + 
\frac{1}{2}\left(\phi^{}_{\PP,\NN}+\phi^{}_{\PP+1,0}\right) \,,
\ee
and the same applies to the equation for $\psi^{}_{\PP,0}$.  In this way, $\psi$ is continuous, to machine
precision, by construction.

Finally, the outgoing (global) boundary conditions described by Eqs.~(\ref{outgoingbcs1}) and~(\ref{outgoingbcs2}) 
are equivalent to the following equations: At the horizon $\rsu=\rsu_{\Hor}$:
\begin{eqnarray}
\partial_{_T}\psi^{}_{1,\Hor} & = & \phi^{}_{1,\Hor}\,, \\
\partial_{_T}\phi^{}_{1,\Hor} & = & \partial_{_T}\varphi^{}_{1,\Hor}\,,\\ 
\partial_{_T}\varphi^{}_{1,\Hor} & = & \left(\partial^{}_{\rsu}\phi^{}_{1}\right)^{}_{\Hor}\,,
\end{eqnarray}
and at spatial infinity $\rsu=\rsu_{\Inf}$:
\begin{eqnarray}
\partial_{_T}\psi^{}_{\DD,\Inf} & = & \phi^{}_{\DD,\Inf}\,, \\ 
\partial_{_T}\phi^{}_{\DD,\Inf} & = &- \partial_{_T}\varphi^{}_{\DD,\Inf}\,, \\
\partial_{_T}\varphi^{}_{\DD,\Inf} & = & \left(\partial^{}_{\rsu} \phi^{}_{\DD}\right)^{}_{\Inf}\,,
\end{eqnarray}
where the subindex $1$ refers to the first subdomain and $D$ to the last one.

%%%%%%%%%%%%%%%%%%%%%%%%%%%%%%%%%%%%%%%%%%%%%%
%   Evolution of the characteristic fields   %
%%%%%%%%%%%%%%%%%%%%%%%%%%%%%%%%%%%%%%%%%%%%%%
\subsection{Evolution of the characteristic fields}\label{charac_evolution}
An alternative to the penalty method is to stick to hyperbolic methods and use
the symmetric hyperbolic structure of our system in order to communicate the
different subdomains by passing the characteristic fields through the 
corresponding characteristics.
This process can be illustrated with the use of Figure~\ref{characteristics}.  
Let us consider the interface between two arbitrary subdomains, say $\Omega_{\Ai}$
and $\Omega_{\Ai+1}$.  The boundary nodes are $r^{\ast}_{\Ai,\regu} = r^{\ast}_{\Ai,\NN}$
and $r^{\ast}_{\Ai+1,\LL}= r^{\ast}_{\Ai+1,0}$.  Then, looking at the representation
of the propagation of the characteristic fields, $U^{\ell m}$ and $V^{\ell, m}$ in
Figure~\ref{characteristics}, we can deduce that at $\rsu = r^{\ast}_{\Ai,\regu}$ we 
only need to evolve Eqs.~(\ref{phys_equ_char}) for $(\psi^{}_{\Ai,\NN},U^{}_{\Ai,\NN})$,
whereas $V^{}_{\Ai,\NN}$ is just obtained by copying the value of $V^{}_{\Ai+1,0}$ from
the other subdomain.  In contrast, at $\rsu = r^{\ast}_{\Ai+1,\LL}$ we only need to evolve
Eqs.~(\ref{phys_equ_char}) for $(\psi^{}_{\Ai+1,0},V^{}_{\Ai+1,0})$,
whereas $Y^{}_{\Ai+1,0}$ is obtained by copying the value of $U^{}_{\Ai,\NN}$ from
the other subdomain.   In the case where the particle is at the interface of these
two subdomains (and hence they are dynamical subdomains), i.e.~$A=P$, we cannot
just copy the values of the characteristic fields, but instead we have to enforce the junction
conditions of Eqs.~(\ref{jump_U}) and~(\ref{jump_V}).  That is, we obtain 
$V^{}_{\PP,\NN}$ by doing:
\be
V^{}_{\PP,\NN} = V^{}_{\PP+1,0} - \saltonum{V}{\PP}{\PP+1}\,, \label{passingV}
\ee
and $U^{}_{\PP+1,0}$ by 
\be
U^{}_{\PP+1,0} = U^{}_{\PP,\NN} + \saltonum{U}{\PP}{\PP+1}\,. \label{passingU}
\ee

The practical implementation of this technique requires to treat the different values of
the fields at the different collocation points in a particular order.  In what follows
we outline an algorithm to do so and the associated equations.  Step by step, the procedure
is the following: (i) For each subdomain $\Omega_{\Ai}$ ($A=1,\ldots,D$), we evolve
the variable $U^{\ell m}$ at the inner points and at the right boundary node, $\rsu =
r^{\ast}_{\Ai,\NN}$, according to the following evolution equation (again we drop the 
harmonic indices $\ell$ and $m$):
\be
\partial_{_T} U^{}_{\Ai,i} = - \left[ 1 + \frac{\left(\partial^{}_{t} X^{}_{\Ai}\right)^{}_{i}}
{\left(\partial^{}_{\rsu} X^{}_{\Ai}\right)^{}_{i}} \right] \left(\partial^{}_{\rsu} U^{}_{\Ai}\right)^{}_{i} 
- V^{\Ai,i}_{\ell}\,\psi^{}_{\Ai,i}\,,
\ee
where $\left(\partial^{}_{\rsu} X^{}_{\Ai}\right)^{}_{i}$ and $\left(\partial^{}_{t} X^{}_{\Ai}\right)^{}_{i}$
are given in Eqs.~(\ref{dX_dr}) and~(\ref{dX_dt}) respectively, and $i=1,\ldots,N$.
The next step is: (ii) For each subdomain $\Omega_{\Ai}$ ($A=1,\ldots,D$), we evolve 
the variable $V^{\ell m}$ at the inner points and at the left boundary node, $\rsu =
r^{\ast}_{\Ai,0}$, according to the following evolution equation:
\be
\partial_{_T} V^{}_{\Ai,i} = \left[ 1 - \frac{\left(\partial^{}_{t} X^{}_{\Ai}\right)^{}_{i}}
{\left(\partial^{}_{\rsu} X^{}_{\Ai}\right)^{}_{i}} \right] \left(\partial^{}_{\rsu} V^{}_{\Ai}\right)^{}_{i} 
- V^{\Ai,i}_{\ell}\,\psi^{}_{\Ai,i}\,,
\ee
where $i=0,\ldots,N-1$.  (iii) We evolve the variable $V^{\ell m}$ at all the right boundary
nodes according to Eq.~(\ref{passingV}).  For the domains without the particle we just simply
pass the value to the right.   Here, we need to make an important remark about the practical
implementation (in a numerical code) of this step.  For the evolution of the time-dependent variables of
our system of ODEs, $\mb{N}_{\Ai,i}(T)$, we use two different methods.  One implements directly
Eq.~(\ref{passingV}).  The other one, uses the method of lines (see, e.g.~\cite{Gustafsson:1995tb}),
so that it incorporates the boundary matching conditions into the evaluation of
the right-hand side of the evolution equations.
Then, to insert this step into the method of lines of our numerical algorithm we need
to use the derivative of Eq.~(\ref{passingV}) rather than this equation itself.  That is,
we need to use:
\be
\frac{d V^{}_{\PP,\NN}}{dT} = \frac{d V^{}_{\PP+1,0}}{d T} - \frac{d \saltonum{V}{\PP}{\PP+1}}{dT}\,.
\ee
This means that we also need the expressions for the derivatives of the jumps in the
variables $U^{\ell m}$ and $V^{\ell m}$. They can be directly derived from the expressions
of the jumps [Eqs.~(\ref{jump_U}) and~(\ref{jump_V})] and the geodesic equations (see
Appendix~\ref{particle_motion}). We give the resulting expressions in Appendix~\ref{jump_derivatives}, in
particular $d\saltonum{V}{\PP}{\PP+1}/dT$ is given in Eq.~(\ref{djumpvdt}).

(iv) We evolve the variable $U^{\ell m}$ at all the left boundary nodes according to Eq.~(\ref{passingU}).
Again, for domains that do not have the particle, we just pass the value of this variable, to the left
in this case.  In order to use the method of lines, following the previous discussion, we need to
use the derivative of Eq.~(\ref{passingU}), that is,
\be
\frac{d U^{}_{\PP+1,0}}{d T} = \frac{d U^{}_{\PP,\NN}}{d T} + \frac{d \saltonum{U}{\PP}{\PP+1}}{dT}\,,
\ee
where $d\saltonum{U}{\PP}{\PP+1}/dT$ is given in Eq.~(\ref{djumpudt}).

(v) We evolve the characteristic variables $U^{\ell m}$ and $V^{\ell m}$ at the global
boundaries.  At the horizon, $\rsu=\rsu_{\Hor}$, based on the method of lines, we use the equations:
\begin{eqnarray}
\partial_{_T}U^{}_{0,\Hor} & = & 0\,,\\
\partial_{_T}V^{}_{0,\Hor} & = & \left(\partial^{}_{\rsu}V^{}_{0}\right)^{}_{\Hor} 
- V^{0,\Hor}_{\ell} \psi^{}_{0,\Hor} \,.
\end{eqnarray}
Then, at spatial infinity $\rsu=\rsu_{\Inf}$:
\begin{eqnarray}
\partial_{_T}U^{}_{\DD,\Inf} & = & -\left(\partial^{}_{\rsu}U^{}_{\DD}\right)^{}_{\Inf}
-V^{\DD,\Inf}_{\ell} \psi^{}_{\DD,\Inf} \,, \\
\partial_{_T}V^{}_{\DD,\Inf} & = & 0\,.
\end{eqnarray}

(vi) The last step consists in evolving the variable $\psi^{\ell m}$ at the collocation 
points of all domains according to the equation
\be
\partial_{_T}\psi^{}_{\Ai,i} = \frac{1}{2}\left(U^{}_{\Ai,i} + V^{}_{\Ai,i}\right)
+ \frac{1}{2}\frac{\left(\partial^{}_{t} X^{}_{\Ai}\right)^{}_{i}}
{\left(\partial^{}_{\rsu} X^{}_{\Ai}\right)^{}_{i}}\left(U^{}_{\Ai,i} - V^{}_{\Ai,i}\right)\,,
\ee
where $i=0,\ldots,N$, and $A=1,\ldots,D$.

%%%%%%%%%%%%%%%%%%%%%%%%%%%%%%%%%%
%  Results from the Simulations  %
%%%%%%%%%%%%%%%%%%%%%%%%%%%%%%%%%%
\section{Results from the Simulations}\label{results}
In this section, we present the main results from the actual numerical calculations.
These calculations have been done using implementations of the PwP formulation in
combination with the multidomain scheme described above.
The details of the numerical codes that we have written for the simulations are essentially the
same as in the case of Paper I, so we refer the reader to this paper for the computational 
details.   Here, we focus on the new aspects of the implementation.

\begin{figure*}
\centering
\includegraphics[width=0.9\textwidth]{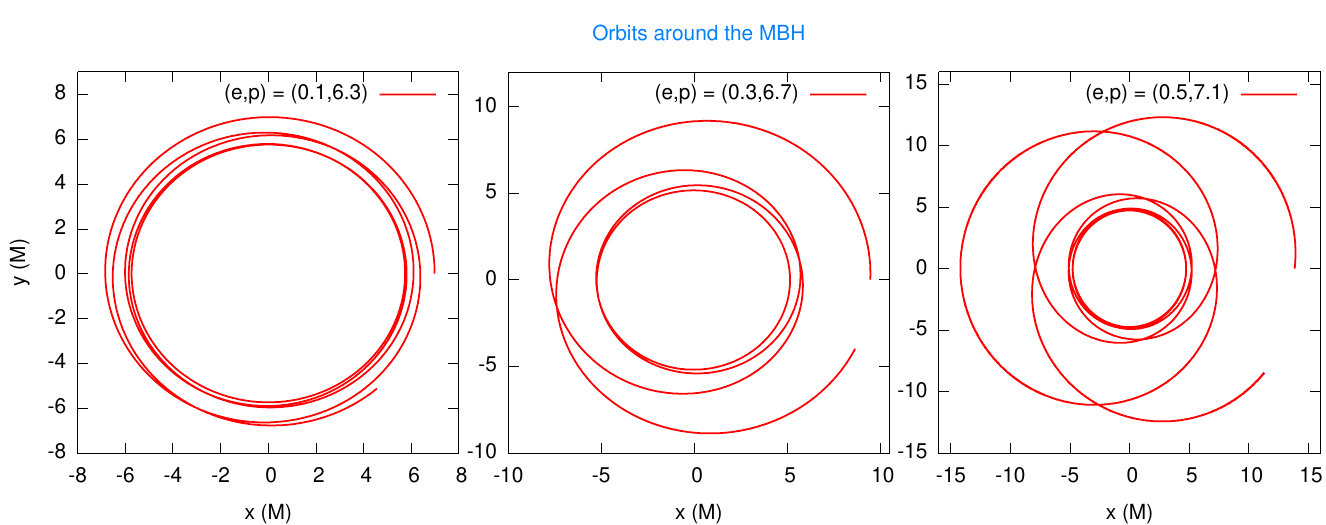}
\caption{Eccentric orbits considered in the numerical evolutions presented in this paper.
From left to right we have: (i) $(e,p)$ $=$ $(0.1,6.3)$; (ii) $(e,p)$ $=$ $(0.3,6.7)$; and 
(iii) $(e,p)$ $=$ $(0.5,7.1)$. These orbits have been integrated using the equations
of Appendix~\ref{particle_motion}, in particular Eqs.~(\ref{chi_dot}) and~(\ref{phi_dot}).}   
\label{orbits}
\end{figure*}

The results that we present refer to three different types of eccentric orbits (see Paper I for
the case of circular orbits).  These three types of orbits, in terms of the eccentricity and
semilatus rectum, are: (i) $(e,p)$ $=$ $(0.1,6.3)$; (ii) $(e,p)$ $=$ $(0.3,6.7)$; and 
(iii) $(e,p)$ $=$ $(0.5,7.1)$.  We show pieces of these orbits in Figure~\ref{orbits}.

%%%%% Figure 3: Orbits
%%%%%

Before introducing the results, it is worth making a comment on the numerical implementation
of the methods and techniques that we described above.   In contrast with the work done for
Paper I, where we only used the penalty method to communicate the subdomains, here we have
implemented in our numerical codes, two different techniques: The penalty method,
described in Sec.~\ref{penalty}, and the direct communication of the characteristic fields, described
in Sec.~\ref{charac_evolution}.  In practice, the numerical implementation of the two methods
is equivalent to having two different numerical codes for the evolution of the system. 
For the results presented in the
paper we have mainly used the penalty formulation.  But we have also used the {\em characteristic}
one, the one based on direct communication of the characteristic fields, both to check the results, 
and for the numerical evaluation of the self-force below.  A last comment on the implementation
refers to the characteristic formulation.  For the time evolution, like in
Paper I, we use a Runge-Kutta 4 (RK4) algorithm (see, e.g.~\cite{Butcher:2008jb,Press:1992nr}).  Then,
as we already discussed in Sec~\ref{charac_evolution}, there are two ways of communicating the
fields: Either we use the expressions containing the jumps in $(U^{\ell m},V^{\ell m})$, Eqs.~(\ref{passingU})
and~(\ref{passingV}), which requires some engineering in the RK4 algorithm, or we use the
expressions containing the derivatives of the jumps in $(U^{\ell m},V^{\ell m})$, Eqs.~(\ref{passingU})
and~(\ref{passingV}), which can be adapted naturally with the method of lines.  This second option requires
to impose initially the values of the jumps, Eqs.~(\ref{djumpudt}) and~(\ref{djumpvdt}), since during
the evolution the only input about them is the information on their derivatives.  Therefore, in this case,
we cannot just use the {\em zero} initial data that we used in Paper I, and that can be used for
the other cases presented here, namely $\mb{U}_{o}$ $=$ $\mb{U}(t=t_{o},r)$
$=$ $(\psi^{\ell m}(t_{o},r), \phi^{\ell m}(t_{o},r), \varphi^{\ell m}(t_{o},r))$ $=$ $\mb{0}$, 
but a modification of it, at least at the nodes where the particle is located.  All these
options have been implemented in our numerical codes.

\begin{figure}[htp]
\centering  
\includegraphics[width=0.44\textwidth]{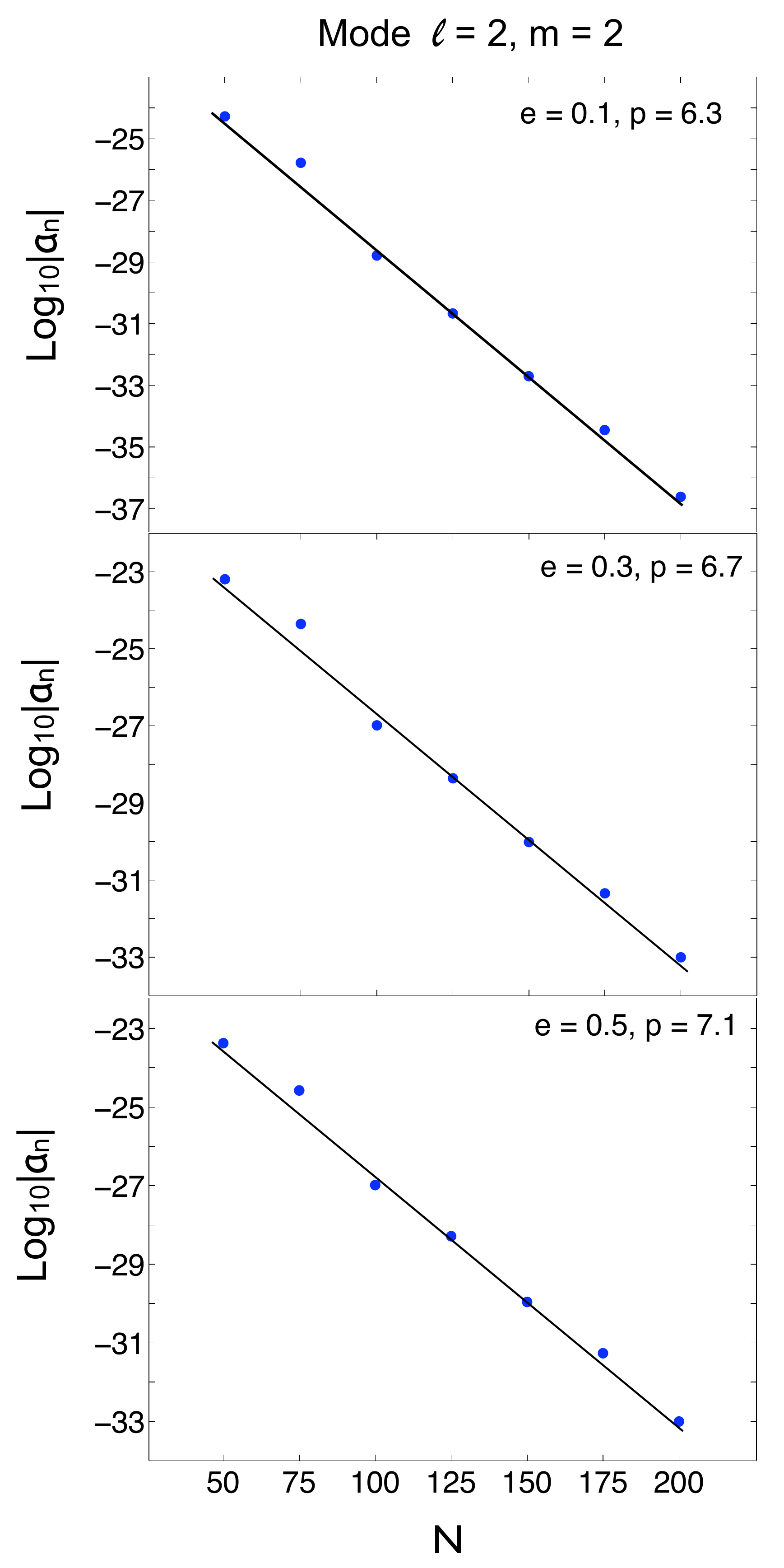}
\caption{We show the dependence of the truncation error, as estimated from the quantity $\log_{10} |a^{}_N|$,
on the number of collocation points, $N$, for evolutions of the harmonic mode $\ell= m = 2$ of the
field variable $\psi^{\ell m}$.  From top to bottom, we show the truncation error for:  (i) $(e,p)$ $=$ $(0.1,6.3)$; 
(ii) $(e,p)$ $=$ $(0.3,6.7)$; and (iii) $(e,p)$ $=$ $(0.5,7.1)$.  The plot corresponds to the solution
on the subdomain to the right of the particle.  The tortoise radial coordinate size of this subdomain,
i.e. $|\rsu_{\PP+1,\NN}-\rsup(t)|$, is in the range $20-40\,M$.  The good fit of the data to a straight
line confirms the exponential convergence of the numerical computations.
\label{convergencewithparticle}}
\end{figure}

The first results that we present refer to the validation of the code.  In Paper I we already
presented several tests.  Here we show a multiple convergence plot (for the different orbits) 
that shows that our PwP scheme indeed provide solutions smooth enough to preserve the \emph{a priori} 
exponential convergence of the PSC method.  In Figure~\ref{convergencewithparticle} we have
plots for the evolution of the harmonic mode $(\ell,m)=(2,2)$, and the different orbits considered
in Figure~\ref{orbits}, showing an estimation of the truncation error, $\log_{10} |a^{}_N|$ [where the
$a^{}_{N}$ is the last spectral coefficient of the spectral approximation of Eq.~(\ref{spectralrepresentation})],
with respect to the number of collocation points, $N$.  For these plots we use the information
coming from the subdomain to the right of the particle, i.e. $\Omega_{\PP+1}$, so that the particle
is located at the left boundary node of this subdomain.  As one can see, the truncation error, 
estimated as the absolute value of the last spectral coefficient, $|a^{}_N|$, decreases exponentially 
with the number of collocation points, as expected in the PSC method for smooth solutions.   

%%%%% Figure 4: Convergence Plot for evolutions with a Particle
%%%%%

\begin{figure*}
\centering 
\includegraphics[width=0.98\textwidth]{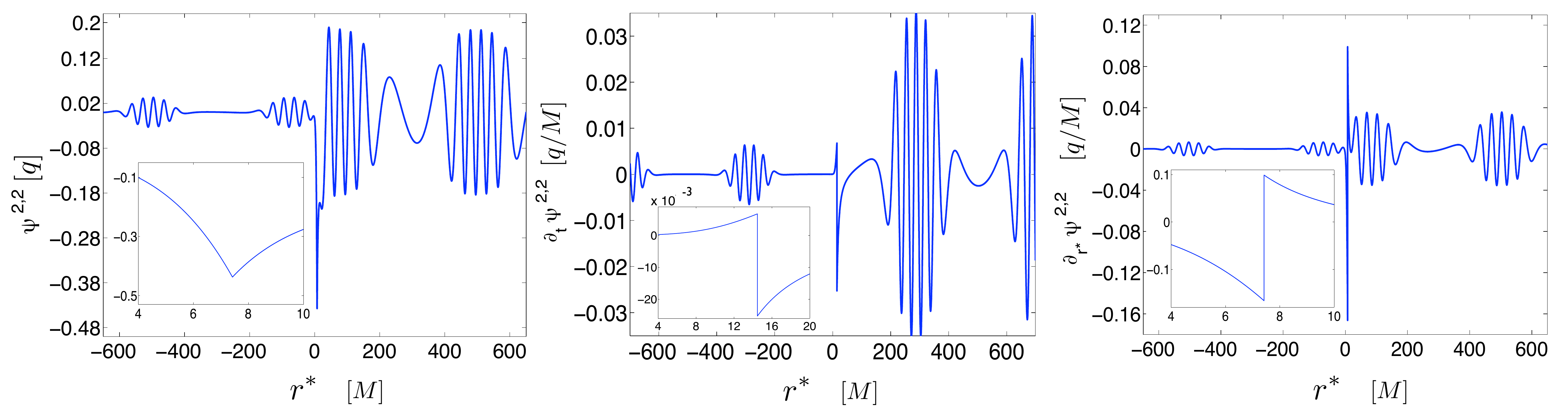}
\caption{Snapshots of the evolution of a scalar charged particle in eccentric orbits around a non-rotating MBH.
The orbital parameters are $(e,p)$ $=$ $(0.5,7.1)$. 
The simulations used $D=10$ subdomains and $N=100$ collocation points per subdomain.  They show the evolution 
of the variables $\psi^{\ell m}$ (left), $\phi^{\ell m}=\partial_t\psi^{\ell m}$ (center), and $\varphi^{\ell m}
=\partial_{\rsu}\psi^{\ell m}$ (right), for the mode $\ell=m=2$.  In particular, they show (see the boxes inside 
the plots) how the jumps in the time and radial derivatives of the  scalar field $\Phi^{\ell m}$ are resolved 
in the multidomain computational framework that implements the PwP scheme.
\label{detailsevolution}}
\end{figure*}

The key ingredient of our PwP scheme is to put the particle at the interface between two 
subdomains, so that it does not appear explicitly as a distributional term in the 
field equations, but as a boundary condition between the solution at adjoining subdomains.
This provides smooth numerical solutions that satisfy the junction conditions dictated by
the jumps [see Eqs.~(\ref{jump_phi}) and~(\ref{jump_varphi}) or Eqs.~(\ref{jump_U}) and~(\ref{jump_V})].
We have performed simulations that use {\em zero initial data} (or initial data that is zero
everywhere except at the particle nodes), that is
\begin{eqnarray}
\psi^{\ell m}(t_o,\rsu) = \phi^{\ell m}(t_o,\rsu) = \varphi^{\ell m}(t_o,\rsu)=0\,.
\label{zeroid}
\end{eqnarray}
As a consequence of this, the numerical evolution produces an initial unphysical burst.  
We have to wait until this unphysical
feature propagates away in order to analyze the solution and obtain physically relevant
results.  In Figure~\ref{detailsevolution} we show snapshots of the evolution of the 
$(\ell,m)=(2,2)$ harmonic mode, including details near the particle 
location, that illustrate the ability of our method to capture the structure
of the solution near the particle, in particular the ability of resolving the jumps in the
time and radial derivative of the field variables.  

%%%%% Figure 5: Details of the Evolution of psi, phi, and varphi
%%%%%

\begin{figure*}
\centering 
\includegraphics[width=0.98\textwidth]{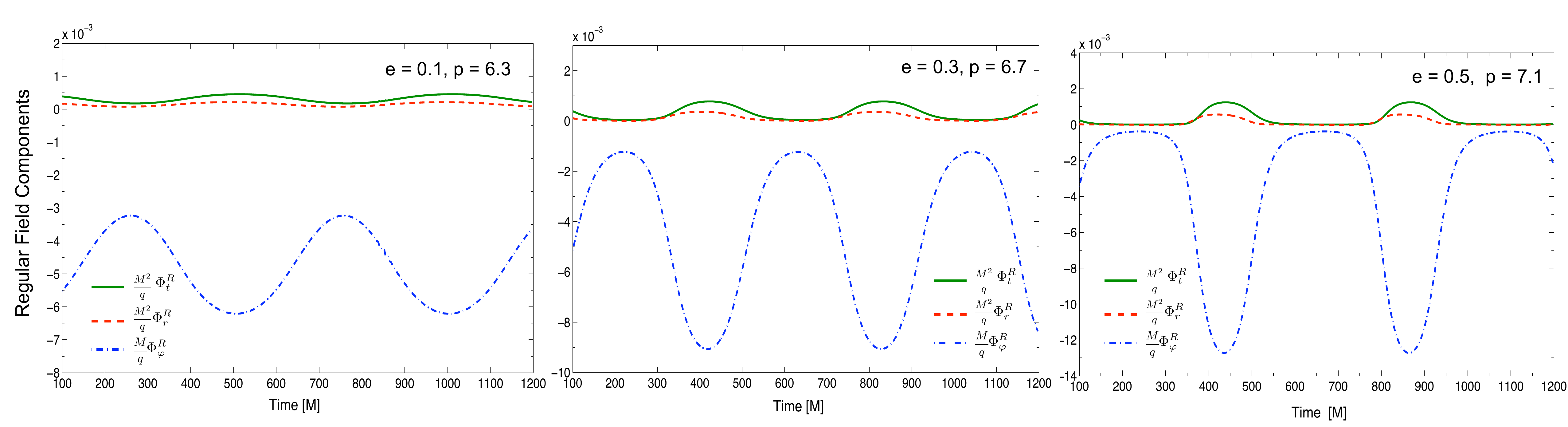}
\caption{Evolution of the components of the gradient of the regularized field, $\nabla^{}_{\alpha}\Phi^{\regu}$,
for a scalar charged particle in eccentric orbits around a non-rotating MBH.  From left to right, the orbital
parameters of the orbits are: (i) $(e,p)$ $=$ $(0.1,6.3)$; (ii) $(e,p)$ $=$ $(0.3,6.7)$; and 
(iii) $(e,p)$ $=$ $(0.5,7.1)$.   For each orbit (frame), the solid line represents the evolution
of the dimensionless time component, $\frac{M^2}{q}\,\Phi^{\regu}_t$; the dashed line represents the evolution
of the dimensionless radial component, $\frac{M^2}{q}\,\Phi^{\regu}_r$; and the dot-dashed line represents the evolution
of the dimensionless azimuthal component, $\frac{M}{q}\,\Phi^{\regu}_\varphi$.
The numerical setup for these calculations used $D=10$ subdomains and 
$N=100$ collocation points per subdomain. 
\label{selfforceevolution}}
\end{figure*}

The main goal of this paper is the computation of the self-force, hence the last two plots
that we present show results on the estimation of the regular field.  From the values of 
the components of the gradient of the regular field, $\nabla^{}_{\alpha}\Phi^{\regu}$, the self-force can be directly 
computed by using Eq.~(\ref{regparticlemotion}).  We have carried out evolutions of the orbits 
shown in Figure~\ref{orbits} and we have computed the values of the components of the gradient
of the regular field along the evolution.  This means that we have evolved all the necessary modes
of the retarded field up to a maximum $\ell$, say $\ell_{\MAX}$.  To compute the total number of
evolutions needed we have to take into account the following facts: (i) We need to compute
the real and imaginary parts [the source term in the field equations~(\ref{master}) is complex as shown
in Eq.~(\ref{source})] for 
each $(\ell,m)$ harmonic component of $\Phi^{\ell m}$; (ii) the real character of the
scalar field implies that for each $(\ell,m)$ the relation 
$\bar\Phi^{\ell\,-m} = (-1)^m\Phi^{\ell m}$ (we do not need to compute the 
modes with $m < 0$) holds; and (iii) For any $(\ell,m)$, we have
\be
Y^{\ell m}(\pi/2,\varphi) = 0~~~\mbox{for $\ell+m$ odd}\,,
\ee
and hence, the jumps for these modes are
exactly zero [see Eq.~(\ref{sourceterm}) and Eqs.~(\ref{jump_phi}),(\ref{jump_varphi}) 
and~(\ref{jump_U}),(\ref{jump_V})] and we do not need to compute them.  
Therefore, the total number of evolutions we need to make in terms of $\ell_{\MAX}$ is
\be
N^{}_{\mbox{\tiny evolutions}} = \frac{\left(\ell_\MAX + 1\right)\left(\ell_\MAX + 2\right)}{2} \,. 
\ee
Once we have evolved
the equations for the retarded field we have to regularize it, $\ell$ by $\ell$, by computing
the components of the gradient of the singular field at the particle location, either using
Eq.~(\ref{singular}) or Eq.~(\ref{singulartetrad}), and subtracting it from the gradient of
the retarded field.  In Figure~\ref{selfforceevolution} we show the results of the computations
of the components of the gradient of the regularized field, which have been obtained by 
choosing $\ell_{\MAX} = 17$, for which we needed to perform $171$ evolutions of the retarded field
equations.  The particular multidomain framework that we employed for these computations 
consists of $D=10$ subdomains and $N=100$ collocation points per subdomain.

%%%%% Figure 6: Evolution of the components of the self-force with time
%%%%%

Finally, we also want to show some numerical values of the components of the gradient of the
regularized field.  To that end, we have chosen to evaluate them at the nearest radial location 
to the pericenter radius [see Eq.~(\ref{periapo})] available in our evolutions.  In 
Table~\ref{self_force_at_pericenter} we show these values for the three orbits of Figure~\ref{orbits}.
In order to have some reference to compare these values, we also provide the values for the last stable 
circular orbit ($r=6\,M$) obtained, using frequency-domain methods, in~\cite{DiazRivera:2004ik}.
Our numerical values have been obtained by using a multidomain framework consisting of $D=80$ subdomains 
and $N=50$ collocation points.  Here, we need to make an important comment: In Paper I, where the
evolutions were always done using the penalty method, we always quoted two values for each component
of the gradient of the regularized field.  The reason is that in our framework we have two different
nodes for the particle (one that belongs to the subdomain to the left of it, $\Omega_{\PP}$, and one
that belongs to the subdomain to the right of it, $\Omega_{\PP+1}$).  Since the equations are 
asymmetric with respect to the particle location, the evolution via penalty method only provides
values of the variables at the particle location that are approximately the same but not identical.
The advantage of this is that it provides a criterium to decide what is the maximum accuracy
of the calculations.  In contrast, the new method of evolution proposed in this paper, the direct
communication of the characteristic fields, provides, by construction, identical values at both
nodes associated with the particle (excepting in the case in which we use spectral filters to
eliminate spurious high-frequency modes in the solution, which is not what we have done for the
results of this paper).  Here, we have presented values obtained with the formulation that uses
the direct communication of characteristic fields (which provides values that are the same,
up to machine precision, at the two particle nodes), but we have checked that they are consistent 
with the values obtained by using the penalty method.

%%%% Table 1: Numerical values of the components of the gradient of the regularized field
\begin{table*}
\begin{minipage}{\textwidth}
\caption{Numerical values of the components of the gradient of the regularized field at the pericenter
radius.  We present our estimations for the orbits shown in Figure~\ref{orbits}: (i) $(e,p)$ $=$ $(0.1,6.3)$
[Column $3$]; (ii) $(e,p)$ $=$ $(0.3,6.7)$ [Column $4$]; and 
(iii) $(e,p)$ $=$ $(0.5,7.1)$ [Column $5$].  For reference, we have included the values for circular
orbits at the last stable circular orbit ($r=6\,M$) obtained, using frequency-domain methods, in~\cite{DiazRivera:2004ik}
[Column $2$].
\label{self_force_at_pericenter}} 
\centering    
\begin{ruledtabular}
\begin{tabular}{l|c|c|c|c}  
    \mbox{}~  & $(e,p) = (0.0,6.0)$ & $(e,p)=(0.1,6.3)$ & $(e,p)=(0.3,6.7)$ & $(e,p)=(0.5,7.1)$ \\[1mm]
\hline
\hline\\[-2mm]
$r^{}_{peri}\; (M)$\protect\footnotemark[1]~ & $6.0$                    & $5.7272727$            & $5.1538461$            & $4.7333333$           \\[2mm]
$r^{}_p\; (M)$\protect\footnotemark[2]       & $6.0$                    & $5.7272925$            & $5.1538801$            & $4.7333989$           \\[2mm]
$\frac{M^2}{q}\,\Phi^{\regu}_t$              & $3.609072\cdot 10^{-4}$  & $4.5171\cdot 10^{-4}$  & $7.6980\cdot 10^{-4}$  & $1.2330\cdot 10^{-3}$ \\[2mm]
$\frac{M^2}{q}\,\Phi^{\regu}_r$              & $1.67728\cdot 10^{-4}$   & $2.1250\cdot 10^{-4}$  & $3.6339\cdot 10^{-4}$  & $5.6122\cdot 10^{-4}$ \\[2mm]
$\frac{M}{q}\,\Phi^{\regu}_\varphi$          & $-5.304231\cdot 10^{-3}$ & $-6.2040\cdot 10^{-3}$ & $-9.0402\cdot 10^{-3}$ & $-1.2685\cdot 10^{-2}$   
\end{tabular} 
\end{ruledtabular}
\footnotetext[1]{This is the exact value of the pericenter radius for each orbit.}
\footnotetext[2]{This is the value of the radial coordinate at which the gradient of the regularized field
has been evaluated.  It is also the nearest value of the radial coordinate to the pericenter value available
in our numerical evolutions.}
\end{minipage}
\end{table*} 
%%%%%

%%%%%%%%%%%%%%%%%%%%%%%%%%%%%%%%
%  CONCLUSIONS AND DISCUSSION  %
%%%%%%%%%%%%%%%%%%%%%%%%%%%%%%%%
\section{Conclusions and Discussion}\label{discussion}

In this paper we have extended the time-domain techniques introduced in Paper I for the
computation of the self-force from the case of circular orbits to generic orbits, what
we have called the PwP formulation.
The key ingredient of this formulation is to consider a multidomain framework in which the
particle is always located at the interface of two subdomains, even in the case of eccentric
orbits.  In this way, the equations that we have solved at each subdomain are homogeneous 
wave-type equations for the retarded field, without distributional sources.  Then, the 
solutions of these equations are sufficiently smooth to provide a good convergence of the numerical
method involved, in our case the PSC method (which provides exponential convergence for smooth solutions).  
The way in which we keep the particle at a fixed node is by using a time-dependent mapping between
the physical domain (the one described by the Schwarzschild radial coordinate $r$) and the
spectral domain (the one in which the calculations of the PSC method take place).  The details
of this mapping are given in Sec.~\ref{pscimplementation}. 
The technical difficulties are then 
transfered to the communication between the different subdomains, in particular between
the subdomains that surround the particle.  In this sense, we have extended
the use of the penalty method to the eccentric case and proposed a new method for the 
communication between subdomains based on the transfer of the characteristic fields of
the hyperbolic system of equations.  These two methods are different enough that their
numerical implementation can be considered to lead to different numerical codes, which is
good for checking the results of the computations.

The main advantages of the PwP formulation in combination with the PSC numerical method 
were already described in Paper I and are still valid for what we have presented in this
paper, so we do not discuss them further here.  Instead, we discuss the possible improvements
of the methods presented.  In Paper I, we already mentioned a number of such improvements:
(i) Compactification of the computational domain, in other to improve the accuracy of the
(global) outgoing boundary conditions which will allow longer evolutions maintaining the
accuracy. (ii) Reduction of the time step, by abandon the linear mapping between the physical
and spectral representations, using a mapping of the type introduced by Kosloff and 
Tal-Ezer~\cite{Kosloff:1993kt}. (iii) Richardson extrapolation, to improve the estimations
of the values of the self-force by using our analytical knowledge of the expansions 
in inverse powers of the harmonic number $\ell$, like in~\cite{Detweiler:2002gi}.
These improvements, although they have not been implemented in the present work, can be
perfectly applied to our framework and have significant potential to improve the efficiency
of the computations.  Another improvement that can be even more important than the ones
already mentioned has to do with the distribution of subdomains.  As it was discussed in 
Paper I, it is very important to choose appropriately the (tortoise coordinate) size 
of the subdomains in order to resolve accurately the different harmonic modes.  In this sense,
the modes with high harmonic number $m$ are more difficult due to the $e^{i m\varphi^{}_{p}(t)}$
dependence of the source term in the original equations [see Eqs.~(\ref{master}) and~(\ref{source})].
This means that the higher $m$ is, the higher the frequency of the scalar field modes generated is, 
and hence the shorter the wavelength gets.  Therefore, these high $m$ modes require more spatial resolution
(also in time).   Both in the calculations of Paper I and in this paper we have used a unique
subdomain structure for all the modes.  Since we must set the size of the subdomains in such a 
way that the high $m$ modes are properly resolved, this means that we are using excessive 
resources for the low $m$ modes.  Thus, the optimal strategy would be to adapt the 
subdomain structure to the type of mode we are dealing with.  One can improve in this way
the efficiency and reduce significantly the computational resources involved.  

Another way of improving the performance of the numerical computations is to provide better
initial data, or at least to reduce the negative effects of the initial burst produced
when we prescribe zero initial data [Eq.~(\ref{zeroid})].  This is important since as it
has been suggested in~\cite{Field:2010xn}, certain components of this unphysical burst
can persist in time, which can deteriorate significantly the precision of the calculations.  
A possible way of alleviating the situation is to {\em switch on} the particle during a certain 
time scale, by multiplying the source terms (in our case the jumps) by a convenient function
of time (starting at zero initially and approaching unity).
We have been exploring this possibility, that turns out to be critical in the
implementation that makes use of Eqs.~(73) and~(74), and we will report on this elsewhere.

To finish, the main goal of the formulation presented in this paper is to develop an
accurate and efficient method to compute the self-force in situations of physical interest.
In particular, for systems of interest for the future observatory LISA.  This means to 
use these techniques in the gravitational case and for spinning MBHs.  In this sense,
we have to mention that while it is not difficult to transfer the techniques discussed
here to the gravitational case, to do the same with the case of a spinning black hole
may require new technical improvements which we will the subject of future investigations.

%%%%%%%%%%%%%%%%%%%%%%
%  ACKNOWLEDGEMENTS  %
%%%%%%%%%%%%%%%%%%%%%%
\acknowledgments 
We would like to thank L. Barack, M. Jasiulek, E. Poisson, L. Rezzolla, U. Sperhake, and B. Wardell 
for helpful discussions and comments. 
PCM is supported by a predoctoral FPU fellowship of the Spanish Ministry of
Science and Innovation (MICINN).
CFS acknowledges support from the Ram\'on y Cajal Programme of the
Spanish Ministry of Education and Science (MEC) and by a Marie Curie
International Reintegration Grant (MIRG-CT-2007-205005/PHY) within the
7th European Community Framework Programme. 
JLJ acknowledges a Fellowship from the Alexander von Humboldt Foundation
and the hospitaly of the Institute of Space Sciences (CSIC-IEEC) during the
realization of this work.
Financial support from the contracts ESP2007-61712 (MEC), FIS2008-06078-C03-01 
and FIS2008-06078-C03-03 (MICINN), and FQM2288 and FQM219 (Junta de Andaluc\'{\i}a) is gratefully 
acknowledged.
This research has used the resources of the Centre de Supercomputaci\'o de Catalunya (CESCA).

%%%%%%%%%%%%%%%%
%  Appendices  %
%%%%%%%%%%%%%%%%
\appendix

%%%%%%%%%%%%%%%%%%%%%
%  Particle motion  %
%%%%%%%%%%%%%%%%%%%%%
\section{Particle motion}\label{particle_motion}

In this section we summarize the form of the geodesic equations for orbits around a Schwarzschild MBH
that we use in this work. Using the constants of motion and the normalization condition of the particle's velocity
$u^{\mu}$, $g_{\mu \nu}u^{\mu}u^{\nu}= -1$, we can reduce the geodesic equations to 
first-order ODEs for $(t^{}_{p},r^{}_{p},\varphi^{}_{p})$.
To avoid the turning points in the integration of the resulting equation for the
radial coordinate, we use an angle variable that is related to the radial coordinate via 
\begin{eqnarray}
r^{}_{p} =\frac{pM}{1+e\cos\chi^{}_{p}}\,,
\end{eqnarray}
where $p$ is the dimensionless semilatus rectum and $e$ is the eccentricity.  From this expression,
the minimum (pericenter, $r^{}_{peri}$) and maximum (apocenter, $r^{}_{apo}$) values of the radial 
coordinate, the turning points, are 
\begin{eqnarray}
r^{}_{peri} =\frac{pM}{1+e}\,, \qquad
r^{}_{apo} =\frac{pM}{1-e}\,. \label{periapo}
\end{eqnarray}
Since the spacetime is static and spherically symmetric, the geodesics have the following
constants of motion: Energy $E^{}_{p}= -u^{}_t$ and angular momentum $L^{}_{p}=u_{\phi}$,
which can be written in terms of the orbital parameters $(p,e)$ as 
\begin{eqnarray}
E^{2}_{p} & = & \frac{(p-2-2e)(p-2+2e)}{p(p-3-e^2)}\,,\\
L^{2}_{p} & = & \frac{p^2M^2}{p-3-e^2}\,.
\end{eqnarray}

The equations for $(\chi^{}_{p}(t),\varphi^{}_{p}(t))$ are 
\begin{widetext}
\begin{eqnarray}
\frac{d\chi^{}_{p}}{dt}&=&\frac{(p-2-2e\cos{\chi^{}_{p}})\sqrt{p-6-2e\cos\chi^{}_{p}}
(1+e\cos\chi^{}_{p})^2}{Mp^2 \sqrt{(p-2)^2-4e^2}}\,, \label{chi_dot}\\[2mm]
\frac{d\varphi^{}_{p}}{dt}&=&\frac{(p-2-2e\cos{\chi^{}_{p}}) (1+e\cos\chi^{}_{p})^2 }
{Mp^{3/2} \sqrt{(p-2)^2-4e^2}}\,. \label{phi_dot}
\end{eqnarray}
\end{widetext}

Two additional expressions of interest for this paper are the first and second time
derivatives of the tortoise coordinate, $\drsup$ and $\ddrsup$.  They can be obtained
from the geodesic equations doing appropriate manipulations.  The result is
\begin{eqnarray}
\drsup = \sqrt{ 1-\frac{f_p}{E_p^2}\left(1+\frac{L_p^2}{r^2_p}\right) } \,,\label{geo_star}
\end{eqnarray}
and
\begin{eqnarray}
\ddrsup = -\frac{f^{}_p}{r^{}_p\, E_p^2}\left\{ \frac{M}{r^{}_p}-\frac{L_p^2}{r_p^2}\left(1-\frac{3M}{r^{}_p}\right)\right\}
\,.\label{dgeo_star}
\end{eqnarray}
%

%%%%%%%%%%%%%%%%%%%%%%%%%%%%%%%%%%%%%%%
%   Structure of the Singular Field   %
%%%%%%%%%%%%%%%%%%%%%%%%%%%%%%%%%%%%%%%
\section{Structure of the Singular Field} \label{singularfield}

In the description of the problem in Sec.~\ref{formulation} it was made clear that the computation
of the self-force requires the regularization of the retarded scalar field.  This can be done in
the framework of the mode-sum scheme, where we need to compute both the harmonic components of the retarded and 
singular parts of the scalar field.  The singular field near the particle has a particular dependence
on the harmonic number $\ell$, which is fully determined by the specification of a set of regularization
parameters (which can be derived by analyzing the field equations in a world-tube close to the particle's
worldline).  In this Appendix we summarize the expressions for these parameters that we have used in this
work (see~\cite{Barack:2002mha,Kim:2004yi,hoon:2005dh,Haas:2006ne}). 

We start 
with the prescription introduced by Haas and Poisson~\cite{Haas:2006ne}, which is the only one 
that includes the regularization parameter that goes with the $\ell^{-2}$ (for large $\ell$)
behaviour of the singular scalar field, namely $D^{}_{\alpha}$.  This term, although it is not
necessary for the computation of the regularized field, it is useful in order to accelerate the convergence 
of it as we increase the value of $\ell^{}_{max}$ (the maximum $\ell$ harmonic number that contributes 
to the computation of the regularized field, and hence to the self-force). 
Following~\cite{Haas:2006ne} we decompose $\nabla^{}_{\nu}\Phi$ in terms of a complex 
tetrad $e^{\alpha}_{(\mu)}(x)$, with the index $(\mu)=\{(0),(+),(-),(3)\}$ labeling the different 
components of this tetrad
\begin{eqnarray}
e^{\alpha}_{(0)}&=& \left[\frac{1}{\sqrt{f}},0,0,0\right]\,,\\
e^{\alpha}_{(3)}&=&\left[0,\sqrt{f}\cos\theta,-\frac{\sin{\theta}}{r},0\right]\,,\\
e^{\alpha}_{(\pm)}&=&\left[0,\sqrt{f}\sin{\theta}e^{\pm i\varphi}, \frac{\cos{\theta}e^{\pm i\varphi}}{r}, 
\frac{\pm ie^{\pm i\varphi}}{r\sin{\theta}}\right]\,. \label{tetrad} 
\end{eqnarray}
The components of the gradient of the retarded field in this tetrad, 
$\Phi^{}_{(\mu)} \equiv \Phi^{}_{\alpha} e^{\alpha}_{(\mu)}$, 
are scalars. Thus, we can expand $\Phi^{}_{(\mu)}$ in scalar spherical harmonics:
\begin{equation}
\Phi_{(\mu)}(x^\alpha)=\sum_{\ell=0}^{\infty}\sum_{m=-\ell}^{\ell}
\Phi_{(\mu)}^{\ell m}(t,r)Y^{\ell m}(\theta,\varphi)\,.
\end{equation}
Then, the components of the gradient of the regular field [Eq.~(\ref{regular2})] are given by
\begin{eqnarray}
\Phi_{\alpha}^{\regu}=\Phi^{}_{\alpha} -\Phi^{\singu}_{\alpha}=
\left\{\Phi^{}_{(\mu)} - \Phi^{\singu}_{(\mu)} \right\}e^{(\mu)}_{\alpha}  \,,\label{reg}
\end{eqnarray}
where, $\Phi^{}_{(\mu)}$ and $\Phi^{S}_{(\mu)}$, are the tetrad projections of the gradient of the 
retarded and singular fields.  Then, the relation between the tetrad and coordinate components of
the gradient of the regular field can be written in the form:
\begin{eqnarray}
\Phi^{\ell m}_{(0)} &=& \frac{1}{\sqrt{f}}\partial_t\Phi^{\ell m}\,,\\
\Phi^{\ell m}_{(+)} &=& -\hat{\alpha}^{\ell m}_{(+)}\left(\Phi^{\ell-1\,m-1}\right) 
                        +\hat{\beta}^{\ell m}_{(+)}\left(\Phi^{\ell+1\,m-1}\right)\,,\\
\Phi^{\ell m}_{(-)} &=& \hat{\alpha}^{\ell m}_{(-)}\left(\Phi^{\ell-1\,m+1}\right)  
                        - \hat{\beta}^{\ell m}_{(-)}\left(\Phi^{\ell+1\,m+1}\right) \,,\\        
\Phi^{\ell m}_{(3)} &=& \hat{\alpha}^{\ell m}_3\left(\Phi^{\ell-1\,m}\right)  
                         +\hat{\beta}^{\ell m}_3\left(\Phi^{\ell+1\,m}\right) \,,                   
\end{eqnarray}
where $\hat{\alpha}^{\ell m}_{(+)}$, $\hat{\beta}^{\ell m}_{(+)}$, $\hat{\alpha}^{\ell m}_{(-)}$, 
$\hat{\beta}^{\ell m}_{(-)}$, $\hat{\alpha}^{\ell m}_3$, and $\hat{\beta}^{\ell m}_3$ are linear
operators given by the following expressions
\begin{eqnarray}
\hat{\alpha}^{\ell m}_{(+)} =\sqrt{\frac{ (\ell +m-1)(\ell +m)}{(2\ell-1)(2\ell+1)} }
\left( \sqrt{f}\frac{\partial}{\partial r} - \frac{\ell-1}{r} \right) \,,
\end{eqnarray}
\begin{eqnarray}
\hat{\alpha}^{\ell m}_{(-)} =\sqrt{\frac{ (\ell- m-1)(\ell -m)}{(2\ell-1)(2\ell+1)} }
\left( \sqrt{f}\frac{\partial}{\partial r} - \frac{\ell-1}{r} \right)\,,
\end{eqnarray}
\begin{eqnarray}
\hat{\alpha}^{}_3 =\sqrt{\frac{ (\ell -m)(\ell +m)}{(2\ell-1)(2\ell+1)} }
\left( \sqrt{f}\frac{\partial}{\partial r} - \frac{\ell-1}{r} \right)\,, 
\end{eqnarray}
\begin{eqnarray}
\hat{\beta}^{\ell m}_{(+)} =\sqrt{\frac{ (\ell -m+1)(\ell -m+2)}{(2\ell+1)(2\ell+3)} }
\left( \sqrt{f}\frac{\partial}{\partial r} + \frac{\ell+2}{r} \right)\,,
\end{eqnarray}
\begin{eqnarray}
\hat{\beta}^{\ell m}_{(-)} =\sqrt{\frac{ (\ell+m+1)(\ell +m+2)}{(2\ell+1)(2\ell+3)} }
\left( \sqrt{f}\frac{\partial}{\partial r} + \frac{\ell+2}{r} \right) \,,
\end{eqnarray}
\begin{eqnarray}
\hat{\beta}^{}_3 =\sqrt{\frac{ (\ell -m+1)(\ell +m+1)}{(2\ell+1)(2\ell+3)} }
\left( \sqrt{f}\frac{\partial}{\partial r} + \frac{\ell+2}{r} \right)\,,
\end{eqnarray}

On the other hand, the tetrad projections of the singular field can be written 
[in a similar way as in the coordinate case of Eq.~(\ref{singular})] as
\begin{eqnarray}
\Phi^{\singu,\ell}_{(\mu)} & = & q\, \left\{ (\ell +\frac{1}{2})A^{}_{(\mu)} +B^{}_{(\mu)} 
+\frac{C^{}_{(\mu)}}{\ell +\frac{1}{2}} \right. \nn \\
& + & \left. \frac{D^{}_{(\mu)}}{(\ell-\frac{1}{2})(\ell+\frac{3}{2})} + \cdots \right\}\,. 
\label{singulartetrad}
\end{eqnarray}
The components $A^{}_{(\mu)}$ of the regularization parameters are given by
\begin{eqnarray}
A^{}_{(0)} &=& \frac{\epsilon^{}_{p}\,E^{}_{p}\,\dot{r}^{}_{p}}
{f^{3/2}_{p}\left(r^2_{p} + L^2_{p}\right)}\,,\\
A^{}_{(+)} &=&\bar{A}_{(-)} = -\epsilon^{}_{p}\, \frac{E^{}_{p} e^{i\varphi^{}_{p}}}
{\sqrt{f^{}_{p}}\left(r^2_{p} + L^2_{p}\right)}\,,\\
A_{(3)} &=& 0\,,
\end{eqnarray}
where we recall that a dot means differentiation with respect to the coordinate time $t$
and $\epsilon^{}_{p}$ is
\be
\epsilon^{}_{p} = \mbox{sign}(r-r^{}_{p}) = \left\{
\begin{array}{ll}
1 & \mbox{if~} r>r^{}_p \,,\\
  & \\
-1 & \mbox{if~} r<r^{}_p \,.\end{array}
 \right.
\ee
The components $B^{}_{(\mu)}$ of the regularization parameters are
\begin{eqnarray}
B^{}_{(0)} = \frac{E^{2}_{p}\,r^{}_{p}\,\dot{r}^{}_{p}}{2\,\left[f^{}_{p}\,\left(r^2_{p} 
+ L^2_{p}\right)\right]^{3/2}} \left(F^{}_{1/2} -2 F^{}_{-1/2}\right)\,,
\end{eqnarray}
where $F^{}_{1/2}$ and $F^{}_{-1/2}$ are objects defined in terms of hypergeometric 
functions~\cite{Abramowitz:1970as} as
\begin{eqnarray}
F^{}_{1/2} = F(\frac{1}{2},\frac{1}{2};1;\frac{L^2_{p}}{r^2_{p}+L^2_{p}})\,, 
\end{eqnarray}
\begin{eqnarray}
F^{}_{-1/2} =  F(-\frac{1}{2},\frac{1}{2};1;\frac{L^2_{p}}{r^2_{p}+L^2_{p}})\,, 
\end{eqnarray}
and
\begin{eqnarray}
B^{}_{(+)} = \bar{B}^{}_{(-)} =e^{i\varphi^{}_{p}}(\Re[B^{}_{(+)}] - i\Im[B^{}_{(+)}])\,,
\end{eqnarray}
where $\Re(z)$ and $\Im(z)$ denote the real and imaginary parts of the complex number $z$.
In the case of $B^{}_{(+)}$, these parts are given by
\begin{widetext}
\begin{eqnarray}
\Re[B^{}_{(+)}]  =  \frac{E^{2}_{p}\,r^{}_{p}\,\dot{r}^2_{p}}{2\,f^{5/2}_{p}\left(r^2_{p} 
+ L^2_{p}\right)^{3/2}}\left( 2F^{}_{-1/2}-F^{}_{1/2} \right)
+ \frac{1}{2\,r^{}_{p}\sqrt{r^2_{p} + L^2_{p}}}\left(\sqrt{f^{}_{p}}\,F^{}_{-1/2} -2\left(\sqrt{f^{}_{p}}-1\right)F^{}_{1/2}   \right)\,, 
\end{eqnarray}
\begin{eqnarray}
\Im[B^{}_{(+)}]  =  \frac{\left(2-\sqrt{f^{}_{p}}\right)\,E^{}_{p}\,\dot{r}^{}_{p}}
{2\,L^{}_{p}\,f^{3/2}_{p}\sqrt{\left(r^2_{p} + L^2_{p}\right)} } \left(F^{}_{1/2} - F^{}_{-1/2}\right)\,.
\end{eqnarray}
\end{widetext}
Moreover, we have
\begin{eqnarray}
B^{}_{(3)} & = & 0\,.
\end{eqnarray}
and 
\begin{eqnarray}
C^{}_{(\mu)}=0\,.
\end{eqnarray}
Finally, the components $D^{}_{(\mu)}$ of the regularization parameters are
\begin{widetext}
\begin{eqnarray}
D^{}_{(0)} & = & \frac{E^{4}_{p}\,r^3_{p}\,\dot{r}^3_{p}}{16\left[f^{}_{p}\left(r^2_{p}+L^2_{p}\right)\right]^{7/2}}
\left\{ \left(5r^2_{p} -3 L^2_{p}\right)F^{}_{1/2} - 8\left(r^2_{p} - L^2_{p}\right)F^{}_{-1/2} \right\}
+\frac{E^{2}_{p}\,\dot{r}^{}_{p}}{16\,f^{3/2}_{p}\left(r^2_{p}+L^2_{p}\right)^{5/2}}
\left\{ \left(F^{}_{1/2}-F^{}_{-1/2}\right) r^3_{p}\right.\nonumber\\
& + & \left.  Mr^2_{p}\left(16\,F^{}_{1/2}-30\,F^{}_{-1/2}\right) - r^{}_{p}L^2_{p}\left(3\,F^{}_{1/2}-7\,F^{}_{-1/2}\right) 
+ ML^2_{p}\left(42\,F^{}_{1/2}-114\,F^{}_{-1/2}\right)\right.\nonumber\\
& + & \left. \frac{ML^4_{p}}{r^2_{p}}\left(18\,F^{}_{1/2}-104\,F^{}_{-1/2}\right)
-\frac{36ML^6_{p}}{r^4_{p}}F^{}_{-1/2}\right\}\,, 
\end{eqnarray}
\end{widetext}
\begin{eqnarray}
D^{}_{(+)} = \bar{D}^{}_{(-)} = e^{i\varphi^{}_{p}}\left(\Re[D^{}_{(+)}] - i\,\Im[D^{}_{(+)}]\right)\,,
\end{eqnarray}
with
\begin{widetext}
\begin{eqnarray}
\Re[D^{}_{(+)}] & = & \frac{E^{4}_{p}\,r^3_{p}\,\dot{r}^4_{p}}{16\,f^{9/2}_{p}\left(r^2_{p}+L^2_{p}\right)^{7/2}}
\left\{8(r^2_{p}-L^2_{p})F^{}_{-1/2}-(5r^2_{p}-3L^2_{p})F^{}_{1/2} \right\}
+ \frac{E^{2}_{p}\,r^{}_{p}\,\dot{r}^2_{p}}{8\,f^{2}_{p}\left(r^2_{p}+L^2_{p}\right)^{3/2}}\left( F^{}_{1/2}-2\,F^{}_{-1/2} \right) \nn \\
& + & \frac{E^{2}_{p}\,\dot{r}^2_{p}}{16\,f^{5/2}_{p}\left(r^2_{p}+L^2_{p}\right)^{5/2}}
\left\{ 4\left(3r^3_{p}+6Mr^2_{p}-L^2_{p}r^{}_{p}+31ML^2_{p}+\frac{26ML^4_{p}}{r^2_{p}}
+\frac{9ML^6_{p}}{r^4_{p}}\right)F^{}_{-1/2}\right.\nonumber\\
& - & \left. \left(7r^3_{p}+12Mr^2_{p}-L^2_{p}r^{}_{p}+46ML^2_{p}+\frac{18ML^4_{p}}{r^2_{p}}\right)F^{}_{1/2} \right\}\nonumber\\   
& + & \frac{\sqrt{f^{}_{p}}}{16(r^2_{p} + L^2_{p})^{3/2}}\left\{\left(3r^{}_{p}+8M+\frac{L^2_{p}}{r^{}_{p}}
+\frac{26ML^2_{p}}{r^2_{p}}+\frac{22ML^4_{p}}{r^4_{p}} +\frac{8ML^6_{p}}{r^6_{p}}\right)F^{}_{-1/2}\right.\nonumber\\   
& - & \left. \left(7r^{}_{p}+6M+\frac{6L^2_{p}}{r^{}_{p}}+\frac{12ML^2_{p}}{r^2_{p}}+\frac{4ML^4_{p}}{r^4_{p}} \right)F^{}_{1/2} \right\}
\nn \\
& - & \frac{1}{8\sqrt{r^2_{p}+L^2_{p}}}\left\{ \left(\frac{1}{r^{}_{p}}+\frac{2M}{r^2_{p}}-\frac{4ML^2_{p}}{r^4_{p}}\right)F^{}_{-1/2} 
 - \frac{3F^{}_{1/2}}{r^{}_{p}} \right\}\,,
\end{eqnarray}
\begin{eqnarray}
\Im[D^{}_{(+)}] & = & \frac{(\sqrt{f^{}_{p}}-2)E^{3}_{p}\,r^2_{p}\,\dot{r}^3_{p}}{16L^{}_{p}\,f^{7/2}_{p}\left(r^2_{p}+L^2_{p}\right)^{5/2}}
\left\{ (r^2_{p}-7L^2_{p})F^{}_{-1/2}-(r^2_{p}-3L^2_{p})F^{}_{1/2}\right\} \nn \\
& + &\frac{E^{}_{p}\,\dot{r}^{}_{p}}{16L^{}_{p}f^{}_{p}\left(r^2_{p}+L^2_{p}\right)^{3/2}}\left\{ \left(4r^2_{p}+2Mr^{}_{p} 
+ 7L^2_{p} + \frac{10ML^2_{p}}{r^{}_{p}} + \frac{14ML^4_{p}}{r^3_{p}}\right)F^{}_{1/2} \right.\nonumber\\
& - & \left. 2\left(2r^2_{p}+Mr^{}_{p}+5L^2_{p}+\frac{10ML^2_{p}}{r^{}_{p}} + \frac{29ML^4_{p}}{r^3_{p}}
+\frac{14ML^6_{p}}{r^5_{p}} \right)F^{}_{-1/2}\right\}\nonumber \\
& + & \frac{E^{}_{p}\,\dot{r}^{}_{p}}{8L^{}_{p}\,f^{3/2}_{p}\left(r^2_{p}+L^2_{p}\right)^{3/2}}
\left\{2\left(r^2_{p}-Mr^{}_{p}+4L^2_{p}-\frac{5ML^2_{p}}{r^{}_{p}}+\frac{2ML^4_{p}}{r^3_{p}}\right)F^{}_{-1/2}\right. \nonumber \\
&&\left. -\left( 2r^2_{p} -2Mr^{}_{p}+5L^2_{p}-\frac{8ML^2_{p}}{r^{}_{p}}\right)F^{}_{1/2}\right\}\;,
\end{eqnarray}
\end{widetext}
and
\begin{eqnarray}
D^{}_{(3)} = 0\,.
\end{eqnarray}

On the other hand, the regularization parameters associated with the coordinate expression of the 
singular field [see Eq.~(\ref{singular})], given in~\cite{Barack:2002mha,Kim:2004yi,hoon:2005dh}, 
are given by
\begin{eqnarray}
A^{}_{t} = -\epsilon^{}_{p}\frac{E^{}_{p}\,\dot{r}^{}_{p}}{f^{}_{p}\left(r^2_{p} + L^2_{p}\right)}\,,
\end{eqnarray}
\begin{eqnarray}
A^{}_{r} = \epsilon^{}_{p} \frac{E^{}_{p}}{f^{}_{p}\left(r^2_{p}+L^2_{p}\right)}\,,
\end{eqnarray}
\begin{eqnarray}
A^{}_{\varphi} = 0\,,
\end{eqnarray}
\begin{eqnarray}
B^{}_{t} = \frac{E^{2}_{p}\,r^{}_{p}\dot{r}^{}_{p}\left(F^{}_{1/2}-2F^{}_{-1/2}\right)}{2\,f^{}_{p}\left(r^2_{p}+L^2_{p}\right)^{3/2}}\,,
\end{eqnarray}
\begin{eqnarray}
B^{}_{r} = \frac{E^{2}_{p}\,r^{}_{p}\left[(\dot{r}^2_{p}-2f^2_{p})F^{}_{1/2}+(\dot{r}^2_{p}-f^2_{p})F^{}_{-1/2}\right]}
{2\,f^{3}_{p}\left(r^2_{p}+L^2_{p}\right)^{3/2}}\,,
\end{eqnarray}
\begin{eqnarray}
B^{}_{\varphi} = \frac{E^{}_{p}\,r^{}_{p}\dot{r}^{}_{p}\left(F^{}_{1/2}-F^{}_{-1/2}\right)}
{2\,L^{}_{p}\,f^{}_{p}\sqrt{r^2_{p}+L^2_{p}}}\,,
\end{eqnarray}
\begin{eqnarray}
C^{}_{\alpha} = 0\,.
\end{eqnarray}
In this case, the regularization parameters $D^{}_{\alpha}$  have not been computed.

%%%%%%%%%%%%%%%%%%%%%
%  Particle motion  %
%%%%%%%%%%%%%%%%%%%%%
\section{Time Derivatives of the Jumps in the Characteristic Variables} \label{jump_derivatives}

The time derivatives of the jumps of the characteristic fields, are given by the following expressions:
\begin{widetext}
\begin{eqnarray}
\frac{d \salto{U}}{dT} &=& - \frac{4\pi q}{E^{}_p r_p^2}\frac{f^{}_p}{1-\drsup} \left\{ \left[  
\left(1-\frac{4M}{r^{}_p}\right)\drsup  -\frac{r^{}_p\,\ddrsup}{1-\drsup}\right] \Re[Y^{\ell,m}]
+m\frac{L^{}_pf^{}_p}{E^{}_pr^{}_p}\,\Im[Y^{\ell,m}] \right\} \nonumber\\
&+&i\left[ \frac{4\pi q}{E^{}_p r_p^2}\frac{f^{}_p}{1-\drsup} \left\{ \left[  
\left(1-\frac{4M}{r^{}_p}\right)\drsup -\frac{r^{}_p\ddrsup}{1-\drsup}\right]\Im[Y^{\ell,m}]
-m\frac{L^{}_pf^{}_p}{E^{}_pr^{}_p}\,\Re[Y^{\ell,m}] \right\} \right]\,,
\label{djumpudt}
\end{eqnarray}
\begin{eqnarray}
\frac{d \salto{V}}{dT} &=& \frac{4\pi q}{E^{}_p r_p^2}\frac{f^{}_p}{1+\drsup} \left\{ \left[ 
\left(1-\frac{4M}{r^{}_p}\right)\drsup + \frac{r^{}_p\ddrsup}{1+\drsup}\right] \Re[Y^{\ell,m}]
+m\frac{L^{}_pf^{}_p}{E^{}_pr^{}_p}\, \Im[Y^{\ell,m}] \right\} \nonumber\\
&-&i\left[ \frac{4\pi q}{E^{}_p r_p^2}\frac{f^{}_p}{1+\drsup} \left\{ \left[  
\left(1-\frac{4M}{r^{}_p}\right)\drsup -\frac{r^{}_p\ddrsup}{1+\drsup}\right]\Im[Y^{\ell,m}]
-m\frac{L^{}_pf^{}_p}{E^{}_pr^{}_p}\,\Re[Y^{\ell,m}] \right\} \right]\,,
\label{djumpvdt}
\end{eqnarray}
\end{widetext}
where $\Re[Y^{\ell,m}]$ and $\Im[Y^{\ell,m}]$ are the real and imaginary parts of the scalar 
spherical harmonic $Y^{\ell m}$, and $\drsup$ and $\ddrsup$ are given by Eqs.~(\ref{geo_star}) 
and~(\ref{dgeo_star}) respectively.

%%%%%%%%%%%%%%%%%%%
%   BIBLIOGRAPHY  %
%%%%%%%%%%%%%%%%%%%
%\bibliography{references}

\begin{thebibliography}{75}
\expandafter\ifx\csname natexlab\endcsname\relax\def\natexlab#1{#1}\fi
\expandafter\ifx\csname bibnamefont\endcsname\relax
  \def\bibnamefont#1{#1}\fi
\expandafter\ifx\csname bibfnamefont\endcsname\relax
  \def\bibfnamefont#1{#1}\fi
\expandafter\ifx\csname citenamefont\endcsname\relax
  \def\citenamefont#1{#1}\fi
\expandafter\ifx\csname url\endcsname\relax
  \def\url#1{\texttt{#1}}\fi
\expandafter\ifx\csname urlprefix\endcsname\relax\def\urlprefix{URL }\fi
\providecommand{\bibinfo}[2]{#2}
\providecommand{\eprint}[2][]{\url{#2}}

\bibitem[{LIG()}]{LIGO}
\emph{\bibinfo{title}{{Laser Interferometer Gravitational-Wave Observatory
  (LIGO)}}}, \bibinfo{note}{{URL: {\tt www.ligo.caltech.edu}}}.

\bibitem[{GEO()}]{GEO600}
\emph{\bibinfo{title}{{GEO600}}}, \bibinfo{note}{{URL: {\tt www.geo600.org}}}.

\bibitem[{VIR()}]{VIRGO}
\emph{\bibinfo{title}{{VIRGO}}}, \bibinfo{note}{{URL: {\tt
  www.virgo.infn.it}}}.

\bibitem[{LCG()}]{LCGT}
\emph{\bibinfo{title}{{Large Scale Cryogenic Gravitational Wave Telescope
  (LCGT)}}}, \bibinfo{note}{{URL: {\tt gw.icrr.u-tokyo.ac.jp/lcgt}}}.

\bibitem[{LIS()}]{LISA}
\emph{\bibinfo{title}{{Laser Interferometer Space Antenna (LISA)}}},
  \bibinfo{note}{{URLs: {\tt www.esa.int}, {\tt lisa.jpl.nasa.gov}}}.

\bibitem[{Ein()}]{Einsteint}
\emph{\bibinfo{title}{{Einstein Telescope}}}, \bibinfo{note}{{URL: {\tt
  www.et-gw.eu}}}.

\bibitem[{\citenamefont{Lobo and
  Sopuerta}(2009{\natexlab{a}})}]{lobosopuertacqg:2009zz}
\bibinfo{editor}{\bibfnamefont{A.}~\bibnamefont{Lobo}} \bibnamefont{and}
  \bibinfo{editor}{\bibfnamefont{C.~F.} \bibnamefont{Sopuerta}}, eds.,
  \emph{\bibinfo{title}{Laser Interferometer Space Antenna. Proceedings of the
  7th International LISA Symposium}}, vol.~\bibinfo{volume}{26}
  (\bibinfo{publisher}{Classical and Quantum Gravity, Institute of Physics},
  \bibinfo{year}{2009}{\natexlab{a}}).

\bibitem[{\citenamefont{Lobo and
  Sopuerta}(2009{\natexlab{b}})}]{lobosopuertajpcs:2009zz}
\bibinfo{editor}{\bibfnamefont{A.}~\bibnamefont{Lobo}} \bibnamefont{and}
  \bibinfo{editor}{\bibfnamefont{C.~F.} \bibnamefont{Sopuerta}}, eds.,
  \emph{\bibinfo{title}{Laser Interferometer Space Antenna. Proceedings of the
  7th International LISA Symposium}}, vol. \bibinfo{volume}{154}
  (\bibinfo{publisher}{Journal of Physics: Conference Series, Institute of
  Physics}, \bibinfo{year}{2009}{\natexlab{b}}).

\bibitem[{\citenamefont{Amaro-Seoane et~al.}(2007)}]{AmaroSeoane:2007aw}
\bibinfo{author}{\bibfnamefont{P.}~\bibnamefont{Amaro-Seoane}}
  \bibnamefont{et~al.}, \bibinfo{journal}{Class. Quant. Grav.}
  \textbf{\bibinfo{volume}{24}}, \bibinfo{pages}{R113} (\bibinfo{year}{2007}),
  \eprint{astro-ph/0703495}.

\bibitem[{\citenamefont{Finn and Thorne}(2000)}]{Finn:2000sy}
\bibinfo{author}{\bibfnamefont{L.~S.} \bibnamefont{Finn}} \bibnamefont{and}
  \bibinfo{author}{\bibfnamefont{K.~S.} \bibnamefont{Thorne}},
  \bibinfo{journal}{Phys. Rev.} \textbf{\bibinfo{volume}{D62}},
  \bibinfo{pages}{124021} (\bibinfo{year}{2000}), \eprint{gr-qc/0007074}.

\bibitem[{\citenamefont{Mino et~al.}(1997)\citenamefont{Mino, Sasaki, and
  Tanaka}}]{Mino:1997nk}
\bibinfo{author}{\bibfnamefont{Y.}~\bibnamefont{Mino}},
  \bibinfo{author}{\bibfnamefont{M.}~\bibnamefont{Sasaki}}, \bibnamefont{and}
  \bibinfo{author}{\bibfnamefont{T.}~\bibnamefont{Tanaka}},
  \bibinfo{journal}{Phys. Rev.} \textbf{\bibinfo{volume}{D55}},
  \bibinfo{pages}{3457} (\bibinfo{year}{1997}), \eprint{gr-qc/9606018}.

\bibitem[{\citenamefont{Quinn and Wald}(1997)}]{Quinn:1997am}
\bibinfo{author}{\bibfnamefont{T.~C.} \bibnamefont{Quinn}} \bibnamefont{and}
  \bibinfo{author}{\bibfnamefont{R.~M.} \bibnamefont{Wald}},
  \bibinfo{journal}{Phys. Rev.} \textbf{\bibinfo{volume}{D56}},
  \bibinfo{pages}{3381} (\bibinfo{year}{1997}), \eprint{gr-qc/9610053}.

\bibitem[{\citenamefont{Gralla and Wald}(2008)}]{Gralla:2008fg}
\bibinfo{author}{\bibfnamefont{S.~E.} \bibnamefont{Gralla}} \bibnamefont{and}
  \bibinfo{author}{\bibfnamefont{R.~M.} \bibnamefont{Wald}},
  \bibinfo{journal}{Class. Quant. Grav.} \textbf{\bibinfo{volume}{25}},
  \bibinfo{pages}{205009} (\bibinfo{year}{2008}), \eprint{0806.3293}.

\bibitem[{\citenamefont{Pound}(2010{\natexlab{a}})}]{Pound:2009sm}
\bibinfo{author}{\bibfnamefont{A.}~\bibnamefont{Pound}},
  \bibinfo{journal}{Phys. Rev.} \textbf{\bibinfo{volume}{D81}},
  \bibinfo{pages}{024023} (\bibinfo{year}{2010}{\natexlab{a}}),
  \eprint{0907.5197}.

\bibitem[{\citenamefont{Pound}(2010{\natexlab{b}})}]{Pound:2010pj}
\bibinfo{author}{\bibfnamefont{A.}~\bibnamefont{Pound}}
  (\bibinfo{year}{2010}{\natexlab{b}}), \eprint{1003.3954}.

\bibitem[{\citenamefont{Glampedakis}(2005)}]{Glampedakis:2005hs}
\bibinfo{author}{\bibfnamefont{K.}~\bibnamefont{Glampedakis}},
  \bibinfo{journal}{Class. Quant. Grav.} \textbf{\bibinfo{volume}{22}},
  \bibinfo{pages}{S605} (\bibinfo{year}{2005}), \eprint{gr-qc/0509024}.

\bibitem[{\citenamefont{Poisson}(2004)}]{Poisson:2004lr}
\bibinfo{author}{\bibfnamefont{E.}~\bibnamefont{Poisson}},
  \bibinfo{journal}{Living Rev. Relativity} \textbf{\bibinfo{volume}{7}},
  \bibinfo{pages}{6} (\bibinfo{year}{2004}), \eprint{gr-qc/0306052},
  \urlprefix\url{http://www.livingreviews.org/lrr-2004-6}.

\bibitem[{\citenamefont{Barack}(2009)}]{Barack:2009fk}
\bibinfo{author}{\bibfnamefont{L.}~\bibnamefont{Barack}}
  (\bibinfo{year}{2009}), \eprint{0908.1664v2},
  \urlprefix\url{http://arXiv.org/abs/0908.1664v2}.

\bibitem[{\citenamefont{Barack and Cutler}(2004)}]{Barack:2003fp}
\bibinfo{author}{\bibfnamefont{L.}~\bibnamefont{Barack}} \bibnamefont{and}
  \bibinfo{author}{\bibfnamefont{C.}~\bibnamefont{Cutler}},
  \bibinfo{journal}{Phys. Rev.} \textbf{\bibinfo{volume}{69}},
  \bibinfo{pages}{082005} (\bibinfo{year}{2004}), \eprint{gr-qc/0310125}.

\bibitem[{\citenamefont{Babak et~al.}(2007)\citenamefont{Babak, Fang, Gair,
  Glampedakis, and Hughes}}]{Babak:2006uv}
\bibinfo{author}{\bibfnamefont{S.}~\bibnamefont{Babak}},
  \bibinfo{author}{\bibfnamefont{H.}~\bibnamefont{Fang}},
  \bibinfo{author}{\bibfnamefont{J.~R.} \bibnamefont{Gair}},
  \bibinfo{author}{\bibfnamefont{K.}~\bibnamefont{Glampedakis}},
  \bibnamefont{and} \bibinfo{author}{\bibfnamefont{S.~A.}
  \bibnamefont{Hughes}}, \bibinfo{journal}{Phys. Rev.}
  \textbf{\bibinfo{volume}{D75}}, \bibinfo{pages}{024005}
  (\bibinfo{year}{2007}), \eprint{gr-qc/0607007}.

\bibitem[{\citenamefont{Drasco and Hughes}(2006)}]{Drasco:2005kz}
\bibinfo{author}{\bibfnamefont{S.}~\bibnamefont{Drasco}} \bibnamefont{and}
  \bibinfo{author}{\bibfnamefont{S.~A.} \bibnamefont{Hughes}},
  \bibinfo{journal}{Phys. Rev.} \textbf{\bibinfo{volume}{D73}},
  \bibinfo{pages}{024027} (\bibinfo{year}{2006}), \eprint{gr-qc/0509101}.

\bibitem[{\citenamefont{Yunes et~al.}(2010)\citenamefont{Yunes, Buonanno,
  Hughes, Coleman~Miller, and Pan}}]{Yunes:2009ef}
\bibinfo{author}{\bibfnamefont{N.}~\bibnamefont{Yunes}},
  \bibinfo{author}{\bibfnamefont{A.}~\bibnamefont{Buonanno}},
  \bibinfo{author}{\bibfnamefont{S.~A.} \bibnamefont{Hughes}},
  \bibinfo{author}{\bibfnamefont{M.}~\bibnamefont{Coleman~Miller}},
  \bibnamefont{and} \bibinfo{author}{\bibfnamefont{Y.}~\bibnamefont{Pan}},
  \bibinfo{journal}{Phys. Rev. Lett.} \textbf{\bibinfo{volume}{104}},
  \bibinfo{pages}{091102} (\bibinfo{year}{2010}), \eprint{0909.4263}.

\bibitem[{\citenamefont{Gair et~al.}(2004)\citenamefont{Gair, Barack,
  Creighton, Cutler, Larson, Phinney, and Vallisneri}}]{Gair:2004iv}
\bibinfo{author}{\bibfnamefont{J.~R.} \bibnamefont{Gair}},
  \bibinfo{author}{\bibfnamefont{L.}~\bibnamefont{Barack}},
  \bibinfo{author}{\bibfnamefont{T.}~\bibnamefont{Creighton}},
  \bibinfo{author}{\bibfnamefont{C.}~\bibnamefont{Cutler}},
  \bibinfo{author}{\bibfnamefont{S.~L.} \bibnamefont{Larson}},
  \bibinfo{author}{\bibfnamefont{E.~S.} \bibnamefont{Phinney}},
  \bibnamefont{and}
  \bibinfo{author}{\bibfnamefont{M.}~\bibnamefont{Vallisneri}},
  \bibinfo{journal}{Class. Quant. Grav.} \textbf{\bibinfo{volume}{21}},
  \bibinfo{pages}{S1595} (\bibinfo{year}{2004}), \eprint{gr-qc/0405137}.

\bibitem[{\citenamefont{Hopman and Alexander}(2006)}]{Hopman:2006xn}
\bibinfo{author}{\bibfnamefont{C.}~\bibnamefont{Hopman}} \bibnamefont{and}
  \bibinfo{author}{\bibfnamefont{T.}~\bibnamefont{Alexander}},
  \bibinfo{journal}{Astrophys. J.} \textbf{\bibinfo{volume}{645}},
  \bibinfo{pages}{L133} (\bibinfo{year}{2006}), \eprint{astro-ph/0603324}.

\bibitem[{\citenamefont{Schutz}(2009)}]{Schutz:2009zz}
\bibinfo{author}{\bibfnamefont{B.~F.} \bibnamefont{Schutz}},
  \bibinfo{journal}{Class. Quant. Grav.} \textbf{\bibinfo{volume}{26}},
  \bibinfo{pages}{094020} (\bibinfo{year}{2009}).

\bibitem[{\citenamefont{MacLeod and Hogan}(2008)}]{MacLeod:2007jd}
\bibinfo{author}{\bibfnamefont{C.~L.} \bibnamefont{MacLeod}} \bibnamefont{and}
  \bibinfo{author}{\bibfnamefont{C.~J.} \bibnamefont{Hogan}},
  \bibinfo{journal}{Phys. Rev.} \textbf{\bibinfo{volume}{D77}},
  \bibinfo{pages}{043512} (\bibinfo{year}{2008}), \eprint{0712.0618}.

\bibitem[{\citenamefont{Gair}(2009)}]{Gair:2008bx}
\bibinfo{author}{\bibfnamefont{J.~R.} \bibnamefont{Gair}},
  \bibinfo{journal}{Class. Quant. Grav.} \textbf{\bibinfo{volume}{26}},
  \bibinfo{pages}{094034} (\bibinfo{year}{2009}), \eprint{0811.0188}.

\bibitem[{\citenamefont{Collins and Hughes}(2004)}]{Collins:2004ex}
\bibinfo{author}{\bibfnamefont{N.~A.} \bibnamefont{Collins}} \bibnamefont{and}
  \bibinfo{author}{\bibfnamefont{S.~A.} \bibnamefont{Hughes}},
  \bibinfo{journal}{Phys. Rev.} \textbf{\bibinfo{volume}{D69}},
  \bibinfo{pages}{124022} (\bibinfo{year}{2004}), \eprint{gr-qc/0402063}.

\bibitem[{\citenamefont{Glampedakis and Babak}(2006)}]{Glampedakis:2005cf}
\bibinfo{author}{\bibfnamefont{K.}~\bibnamefont{Glampedakis}} \bibnamefont{and}
  \bibinfo{author}{\bibfnamefont{S.}~\bibnamefont{Babak}},
  \bibinfo{journal}{Class. Quant. Grav.} \textbf{\bibinfo{volume}{23}},
  \bibinfo{pages}{4167} (\bibinfo{year}{2006}), \eprint{gr-qc/0510057}.

\bibitem[{\citenamefont{Barack and Cutler}(2007)}]{Barack:2006pq}
\bibinfo{author}{\bibfnamefont{L.}~\bibnamefont{Barack}} \bibnamefont{and}
  \bibinfo{author}{\bibfnamefont{C.}~\bibnamefont{Cutler}},
  \bibinfo{journal}{Phys. Rev.} \textbf{\bibinfo{volume}{D75}},
  \bibinfo{pages}{042003} (\bibinfo{year}{2007}), \eprint{gr-qc/0612029}.

\bibitem[{\citenamefont{Yunes et~al.}(2008)\citenamefont{Yunes, Sopuerta,
  Rubbo, and Holley-Bockelmann}}]{Yunes:2007zp}
\bibinfo{author}{\bibfnamefont{N.}~\bibnamefont{Yunes}},
  \bibinfo{author}{\bibfnamefont{C.~F.} \bibnamefont{Sopuerta}},
  \bibinfo{author}{\bibfnamefont{L.~J.} \bibnamefont{Rubbo}}, \bibnamefont{and}
  \bibinfo{author}{\bibfnamefont{K.}~\bibnamefont{Holley-Bockelmann}},
  \bibinfo{journal}{Astrophys. J.} \textbf{\bibinfo{volume}{675}},
  \bibinfo{pages}{604} (\bibinfo{year}{2008}), \eprint{0704.2612}.

\bibitem[{\citenamefont{Brink}(2008)}]{Brink:2008xx}
\bibinfo{author}{\bibfnamefont{J.}~\bibnamefont{Brink}},
  \bibinfo{journal}{Phys. Rev.} \textbf{\bibinfo{volume}{D78}},
  \bibinfo{pages}{102001} (\bibinfo{year}{2008}), \eprint{0807.1178}.

\bibitem[{\citenamefont{Vigeland and Hughes}(2010)}]{Vigeland:2009pr}
\bibinfo{author}{\bibfnamefont{S.~J.} \bibnamefont{Vigeland}} \bibnamefont{and}
  \bibinfo{author}{\bibfnamefont{S.~A.} \bibnamefont{Hughes}},
  \bibinfo{journal}{Phys. Rev.} \textbf{\bibinfo{volume}{D81}},
  \bibinfo{pages}{024030} (\bibinfo{year}{2010}), \eprint{0911.1756}.

\bibitem[{\citenamefont{Hughes}(2010)}]{Hughes:2010xf}
\bibinfo{author}{\bibfnamefont{S.~A.} \bibnamefont{Hughes}}
  (\bibinfo{year}{2010}), \eprint{1002.2591}.

\bibitem[{\citenamefont{Hughes}(2006)}]{Hughes:2006pm}
\bibinfo{author}{\bibfnamefont{S.~A.} \bibnamefont{Hughes}},
  \bibinfo{journal}{AIP Conf. Proc.} \textbf{\bibinfo{volume}{873}},
  \bibinfo{pages}{233} (\bibinfo{year}{2006}), \eprint{gr-qc/0608140}.

\bibitem[{\citenamefont{Sopuerta and Yunes}(2009)}]{Sopuerta:2009iy}
\bibinfo{author}{\bibfnamefont{C.~F.} \bibnamefont{Sopuerta}} \bibnamefont{and}
  \bibinfo{author}{\bibfnamefont{N.}~\bibnamefont{Yunes}},
  \bibinfo{journal}{Phys. Rev.} \textbf{\bibinfo{volume}{D80}},
  \bibinfo{pages}{064006} (\bibinfo{year}{2009}), \eprint{0904.4501}.

\bibitem[{\citenamefont{Miller and Colbert}(2004)}]{Miller:2003sc}
\bibinfo{author}{\bibfnamefont{M.~C.} \bibnamefont{Miller}} \bibnamefont{and}
  \bibinfo{author}{\bibfnamefont{E.~J.~M.} \bibnamefont{Colbert}},
  \bibinfo{journal}{Int. J. Mod. Phys.} \textbf{\bibinfo{volume}{D13}},
  \bibinfo{pages}{1} (\bibinfo{year}{2004}), \eprint{astro-ph/0308402}.

\bibitem[{\citenamefont{Miller}(2008)}]{Miller:2008fi}
\bibinfo{author}{\bibfnamefont{M.~C.} \bibnamefont{Miller}}
  (\bibinfo{year}{2008}), \eprint{0812.3028}.

\bibitem[{\citenamefont{Brown et~al.}(2007)}]{Brown06}
\bibinfo{author}{\bibfnamefont{D.~A.} \bibnamefont{Brown}}
  \bibnamefont{et~al.}, \bibinfo{journal}{Phys. Rev. Lett.}
  \textbf{\bibinfo{volume}{99}}, \bibinfo{pages}{201102}
  (\bibinfo{year}{2007}), \eprint{gr-qc/0612060}.

\bibitem[{\citenamefont{Mandel et~al.}(2007)\citenamefont{Mandel, Brown, Gair,
  and Miller}}]{Mandel:2007hi}
\bibinfo{author}{\bibfnamefont{I.}~\bibnamefont{Mandel}},
  \bibinfo{author}{\bibfnamefont{D.~A.} \bibnamefont{Brown}},
  \bibinfo{author}{\bibfnamefont{J.~R.} \bibnamefont{Gair}}, \bibnamefont{and}
  \bibinfo{author}{\bibfnamefont{M.~C.} \bibnamefont{Miller}}
  (\bibinfo{year}{2007}), \eprint{0705.0285}.

\bibitem[{\citenamefont{Barack and Ori}(2000)}]{Barack:1999wf}
\bibinfo{author}{\bibfnamefont{L.}~\bibnamefont{Barack}} \bibnamefont{and}
  \bibinfo{author}{\bibfnamefont{A.}~\bibnamefont{Ori}},
  \bibinfo{journal}{Phys. Rev.} \textbf{\bibinfo{volume}{D61}},
  \bibinfo{pages}{061502} (\bibinfo{year}{2000}), \eprint{gr-qc/9912010}.

\bibitem[{\citenamefont{Barack}(2000)}]{Barack:2000eh}
\bibinfo{author}{\bibfnamefont{L.}~\bibnamefont{Barack}},
  \bibinfo{journal}{Phys. Rev.} \textbf{\bibinfo{volume}{D62}},
  \bibinfo{pages}{084027} (\bibinfo{year}{2000}), \eprint{gr-qc/0005042}.

\bibitem[{\citenamefont{Barack}(2001)}]{Barack:2001bw}
\bibinfo{author}{\bibfnamefont{L.}~\bibnamefont{Barack}},
  \bibinfo{journal}{Phys. Rev.} \textbf{\bibinfo{volume}{D64}},
  \bibinfo{pages}{084021} (\bibinfo{year}{2001}), \eprint{gr-qc/0105040}.

\bibitem[{\citenamefont{Mino et~al.}(2002)\citenamefont{Mino, Nakano, and
  Sasaki}}]{Mino:2001mq}
\bibinfo{author}{\bibfnamefont{Y.}~\bibnamefont{Mino}},
  \bibinfo{author}{\bibfnamefont{H.}~\bibnamefont{Nakano}}, \bibnamefont{and}
  \bibinfo{author}{\bibfnamefont{M.}~\bibnamefont{Sasaki}},
  \bibinfo{journal}{Prog. Theor. Phys.} \textbf{\bibinfo{volume}{108}},
  \bibinfo{pages}{1039} (\bibinfo{year}{2002}), \eprint{gr-qc/0111074}.

\bibitem[{\citenamefont{Barack et~al.}(2002)\citenamefont{Barack, Mino, Nakano,
  Ori, and Sasaki}}]{Barack:2001gx}
\bibinfo{author}{\bibfnamefont{L.}~\bibnamefont{Barack}},
  \bibinfo{author}{\bibfnamefont{Y.}~\bibnamefont{Mino}},
  \bibinfo{author}{\bibfnamefont{H.}~\bibnamefont{Nakano}},
  \bibinfo{author}{\bibfnamefont{A.}~\bibnamefont{Ori}}, \bibnamefont{and}
  \bibinfo{author}{\bibfnamefont{M.}~\bibnamefont{Sasaki}},
  \bibinfo{journal}{Phys. Rev. Lett.} \textbf{\bibinfo{volume}{88}},
  \bibinfo{pages}{091101} (\bibinfo{year}{2002}), \eprint{gr-qc/0111001}.

\bibitem[{\citenamefont{Barack and Ori}(2002)}]{Barack:2002mha}
\bibinfo{author}{\bibfnamefont{L.}~\bibnamefont{Barack}} \bibnamefont{and}
  \bibinfo{author}{\bibfnamefont{A.}~\bibnamefont{Ori}},
  \bibinfo{journal}{Phys. Rev.} \textbf{\bibinfo{volume}{D66}},
  \bibinfo{pages}{084022} (\bibinfo{year}{2002}), \eprint{gr-qc/0204093}.

\bibitem[{\citenamefont{Detweiler et~al.}(2003)\citenamefont{Detweiler,
  Messaritaki, and Whiting}}]{Detweiler:2002gi}
\bibinfo{author}{\bibfnamefont{S.}~\bibnamefont{Detweiler}},
  \bibinfo{author}{\bibfnamefont{E.}~\bibnamefont{Messaritaki}},
  \bibnamefont{and} \bibinfo{author}{\bibfnamefont{B.~F.}
  \bibnamefont{Whiting}}, \bibinfo{journal}{Phys. Rev.}
  \textbf{\bibinfo{volume}{D67}}, \bibinfo{pages}{104016}
  (\bibinfo{year}{2003}), \eprint{gr-qc/0205079}.

\bibitem[{\citenamefont{Haas and Poisson}(2006)}]{Haas:2006ne}
\bibinfo{author}{\bibfnamefont{R.}~\bibnamefont{Haas}} \bibnamefont{and}
  \bibinfo{author}{\bibfnamefont{E.}~\bibnamefont{Poisson}},
  \bibinfo{journal}{Phys. Rev.} \textbf{\bibinfo{volume}{D74}},
  \bibinfo{pages}{044009} (\bibinfo{year}{2006}), \eprint{gr-qc/0605077}.

\bibitem[{\citenamefont{Sopuerta et~al.}(2006)\citenamefont{Sopuerta, Sun,
  Laguna, and Xu}}]{Sopuerta:2005rd}
\bibinfo{author}{\bibfnamefont{C.~F.} \bibnamefont{Sopuerta}},
  \bibinfo{author}{\bibfnamefont{P.}~\bibnamefont{Sun}},
  \bibinfo{author}{\bibfnamefont{P.}~\bibnamefont{Laguna}}, \bibnamefont{and}
  \bibinfo{author}{\bibfnamefont{J.}~\bibnamefont{Xu}},
  \bibinfo{journal}{Class. Quantum Grav.} \textbf{\bibinfo{volume}{23}},
  \bibinfo{pages}{251} (\bibinfo{year}{2006}), \eprint{gr-qc/0507112}.

\bibitem[{\citenamefont{Barack and Golbourn}(2007)}]{Barack:2007jh}
\bibinfo{author}{\bibfnamefont{L.}~\bibnamefont{Barack}} \bibnamefont{and}
  \bibinfo{author}{\bibfnamefont{D.~A.} \bibnamefont{Golbourn}},
  \bibinfo{journal}{Phys. Rev.} \textbf{\bibinfo{volume}{D76}},
  \bibinfo{pages}{044020} (\bibinfo{year}{2007}), \eprint{0705.3620}.

\bibitem[{\citenamefont{Barack et~al.}(2007)\citenamefont{Barack, Golbourn, and
  Sago}}]{Barack:2007we}
\bibinfo{author}{\bibfnamefont{L.}~\bibnamefont{Barack}},
  \bibinfo{author}{\bibfnamefont{D.~A.} \bibnamefont{Golbourn}},
  \bibnamefont{and} \bibinfo{author}{\bibfnamefont{N.}~\bibnamefont{Sago}},
  \bibinfo{journal}{Phys. Rev.} \textbf{\bibinfo{volume}{D76}},
  \bibinfo{pages}{124036} (\bibinfo{year}{2007}), \eprint{0709.4588}.

\bibitem[{\citenamefont{Vega and Detweiler}(2008)}]{Vega:2007mc}
\bibinfo{author}{\bibfnamefont{I.}~\bibnamefont{Vega}} \bibnamefont{and}
  \bibinfo{author}{\bibfnamefont{S.}~\bibnamefont{Detweiler}},
  \bibinfo{journal}{Phys. Rev.} \textbf{\bibinfo{volume}{D77}},
  \bibinfo{pages}{084008} (\bibinfo{year}{2008}), \eprint{0712.4405}.

\bibitem[{\citenamefont{Lousto and Nakano}(2008)}]{Lousto:2008mb}
\bibinfo{author}{\bibfnamefont{C.~O.} \bibnamefont{Lousto}} \bibnamefont{and}
  \bibinfo{author}{\bibfnamefont{H.}~\bibnamefont{Nakano}},
  \bibinfo{journal}{Class. Quant. Grav.} \textbf{\bibinfo{volume}{25}},
  \bibinfo{pages}{145018} (\bibinfo{year}{2008}), \eprint{0802.4277}.

\bibitem[{\citenamefont{Brandt and Bruegmann}(1997)}]{Brandt:1997tf}
\bibinfo{author}{\bibfnamefont{S.}~\bibnamefont{Brandt}} \bibnamefont{and}
  \bibinfo{author}{\bibfnamefont{B.}~\bibnamefont{Bruegmann}},
  \bibinfo{journal}{Phys. Rev. Lett.} \textbf{\bibinfo{volume}{78}},
  \bibinfo{pages}{3606} (\bibinfo{year}{1997}), \eprint{gr-qc/9703066}.

\bibitem[{\citenamefont{Campanelli et~al.}(2006)\citenamefont{Campanelli,
  Lousto, Marronetti, and Zlochower}}]{Campanelli:2005dd}
\bibinfo{author}{\bibfnamefont{M.}~\bibnamefont{Campanelli}},
  \bibinfo{author}{\bibfnamefont{C.~O.} \bibnamefont{Lousto}},
  \bibinfo{author}{\bibfnamefont{P.}~\bibnamefont{Marronetti}},
  \bibnamefont{and}
  \bibinfo{author}{\bibfnamefont{Y.}~\bibnamefont{Zlochower}},
  \bibinfo{journal}{Phys. Rev. Lett.} \textbf{\bibinfo{volume}{96}},
  \bibinfo{pages}{111101} (\bibinfo{year}{2006}), \eprint{gr-qc/0511048}.

\bibitem[{\citenamefont{Baker et~al.}(2006)\citenamefont{Baker, Centrella,
  Choi, Koppitz, and van Meter}}]{Baker:2005vv}
\bibinfo{author}{\bibfnamefont{J.~G.} \bibnamefont{Baker}},
  \bibinfo{author}{\bibfnamefont{J.}~\bibnamefont{Centrella}},
  \bibinfo{author}{\bibfnamefont{D.-I.} \bibnamefont{Choi}},
  \bibinfo{author}{\bibfnamefont{M.}~\bibnamefont{Koppitz}}, \bibnamefont{and}
  \bibinfo{author}{\bibfnamefont{J.}~\bibnamefont{van Meter}},
  \bibinfo{journal}{Phys. Rev. Lett.} \textbf{\bibinfo{volume}{96}},
  \bibinfo{pages}{111102} (\bibinfo{year}{2006}), \eprint{gr-qc/0511103}.

\bibitem[{\citenamefont{Canizares and
  Sopuerta}(2009{\natexlab{a}})}]{Canizares:2008dp}
\bibinfo{author}{\bibfnamefont{P.}~\bibnamefont{Canizares}} \bibnamefont{and}
  \bibinfo{author}{\bibfnamefont{C.~F.} \bibnamefont{Sopuerta}},
  \bibinfo{journal}{J. Phys. Conf. Ser.} \textbf{\bibinfo{volume}{154}},
  \bibinfo{pages}{012053} (\bibinfo{year}{2009}{\natexlab{a}}),
  \eprint{0811.0294}.

\bibitem[{\citenamefont{Canizares and
  Sopuerta}(2009{\natexlab{b}})}]{Canizares:2009ay}
\bibinfo{author}{\bibfnamefont{P.}~\bibnamefont{Canizares}} \bibnamefont{and}
  \bibinfo{author}{\bibfnamefont{C.~F.} \bibnamefont{Sopuerta}},
  \bibinfo{journal}{Phys. Rev.} \textbf{\bibinfo{volume}{D79}},
  \bibinfo{pages}{084020} (\bibinfo{year}{2009}{\natexlab{b}}),
  \eprint{0903.0505}.

\bibitem[{\citenamefont{Field et~al.}(2009)\citenamefont{Field, Hesthaven, and
  Lau}}]{Field:2009kk}
\bibinfo{author}{\bibfnamefont{S.~E.} \bibnamefont{Field}},
  \bibinfo{author}{\bibfnamefont{J.~S.} \bibnamefont{Hesthaven}},
  \bibnamefont{and} \bibinfo{author}{\bibfnamefont{S.~R.} \bibnamefont{Lau}},
  \bibinfo{journal}{Class. Quant. Grav.} \textbf{\bibinfo{volume}{26}},
  \bibinfo{pages}{165010} (\bibinfo{year}{2009}), \eprint{0902.1287}.

\bibitem[{\citenamefont{Haas}(2007)}]{Haas:2007kz}
\bibinfo{author}{\bibfnamefont{R.}~\bibnamefont{Haas}}, \bibinfo{journal}{Phys.
  Rev.} \textbf{\bibinfo{volume}{D75}}, \bibinfo{pages}{124011}
  (\bibinfo{year}{2007}), \eprint{0704.0797}.

\bibitem[{\citenamefont{Diaz-Rivera et~al.}(2004)\citenamefont{Diaz-Rivera,
  Messaritaki, Whiting, and Detweiler}}]{DiazRivera:2004ik}
\bibinfo{author}{\bibfnamefont{L.~M.} \bibnamefont{Diaz-Rivera}},
  \bibinfo{author}{\bibfnamefont{E.}~\bibnamefont{Messaritaki}},
  \bibinfo{author}{\bibfnamefont{B.~F.} \bibnamefont{Whiting}},
  \bibnamefont{and}
  \bibinfo{author}{\bibfnamefont{S.}~\bibnamefont{Detweiler}},
  \bibinfo{journal}{Phys. Rev.} \textbf{\bibinfo{volume}{D70}},
  \bibinfo{pages}{124018} (\bibinfo{year}{2004}), \eprint{gr-qc/0410011}.

\bibitem[{\citenamefont{Detweiler and Whiting}(2003)}]{Detweiler:2002mi}
\bibinfo{author}{\bibfnamefont{S.}~\bibnamefont{Detweiler}} \bibnamefont{and}
  \bibinfo{author}{\bibfnamefont{B.~F.} \bibnamefont{Whiting}},
  \bibinfo{journal}{Phys. Rev.} \textbf{\bibinfo{volume}{D67}},
  \bibinfo{pages}{024025} (\bibinfo{year}{2003}), \eprint{gr-qc/0202086}.

\bibitem[{\citenamefont{Kim}(2004)}]{Kim:2004yi}
\bibinfo{author}{\bibfnamefont{D.-H.} \bibnamefont{Kim}}
  (\bibinfo{year}{2004}), \eprint{gr-qc/0402014}.

\bibitem[{\citenamefont{Kim}(2005)}]{hoon:2005dh}
\bibinfo{author}{\bibfnamefont{D.-H.} \bibnamefont{Kim}}, Ph.D. thesis,
  \bibinfo{school}{University of Florida} (\bibinfo{year}{2005}).

\bibitem[{\citenamefont{Courant and Hilbert}(1966)}]{Hilbert:1966ch}
\bibinfo{author}{\bibfnamefont{R.}~\bibnamefont{Courant}} \bibnamefont{and}
  \bibinfo{author}{\bibfnamefont{D.}~\bibnamefont{Hilbert}},
  \emph{\bibinfo{title}{Methods of Mathematical Physics}}, vol.
  \bibinfo{volume}{II. Partial Differential Equations}
  (\bibinfo{publisher}{Interscience Publishers, John Wiley \& Sons},
  \bibinfo{address}{New York}, \bibinfo{year}{1966}).

\bibitem[{\citenamefont{Sopuerta and Laguna}(2006)}]{Sopuerta:2005gz}
\bibinfo{author}{\bibfnamefont{C.~F.} \bibnamefont{Sopuerta}} \bibnamefont{and}
  \bibinfo{author}{\bibfnamefont{P.}~\bibnamefont{Laguna}},
  \bibinfo{journal}{Phys. Rev.} \textbf{\bibinfo{volume}{D73}},
  \bibinfo{pages}{044028} (\bibinfo{year}{2006}), \eprint{gr-qc/0512028}.

\bibitem[{\citenamefont{Pfeiffer et~al.}(2003)\citenamefont{Pfeiffer, Kidder,
  Scheel, and Teukolsky}}]{Pfeiffer:2002wt}
\bibinfo{author}{\bibfnamefont{H.~P.} \bibnamefont{Pfeiffer}},
  \bibinfo{author}{\bibfnamefont{L.~E.} \bibnamefont{Kidder}},
  \bibinfo{author}{\bibfnamefont{M.~A.} \bibnamefont{Scheel}},
  \bibnamefont{and} \bibinfo{author}{\bibfnamefont{S.~A.}
  \bibnamefont{Teukolsky}}, \bibinfo{journal}{Comput. Phys. Commun.}
  \textbf{\bibinfo{volume}{152}}, \bibinfo{pages}{253} (\bibinfo{year}{2003}),
  \eprint{gr-qc/0202096}.

\bibitem[{\citenamefont{Boyd}(2001)}]{Boyd}
\bibinfo{author}{\bibfnamefont{J.~P.} \bibnamefont{Boyd}},
  \emph{\bibinfo{title}{Chebyshev and Fourier Spectral Methods}}
  (\bibinfo{publisher}{Dover}, \bibinfo{address}{New York},
  \bibinfo{year}{2001}), \bibinfo{edition}{2nd} ed.

\bibitem[{\citenamefont{Hesthaven}(2000)}]{Hesthaven:2000jh}
\bibinfo{author}{\bibfnamefont{J.~S.} \bibnamefont{Hesthaven}},
  \bibinfo{journal}{Applied Numerical Mathematics}
  \textbf{\bibinfo{volume}{33}}, \bibinfo{pages}{23} (\bibinfo{year}{2000}).

\bibitem[{\citenamefont{Gustafsson et~al.}(1995)\citenamefont{Gustafsson,
  Kreiss, and Oliger}}]{Gustafsson:1995tb}
\bibinfo{author}{\bibfnamefont{B.}~\bibnamefont{Gustafsson}},
  \bibinfo{author}{\bibfnamefont{H.}~\bibnamefont{Kreiss}}, \bibnamefont{and}
  \bibinfo{author}{\bibfnamefont{J.}~\bibnamefont{Oliger}},
  \emph{\bibinfo{title}{Time dependent problems}} (\bibinfo{publisher}{John
  Wiley \& Sons}, \bibinfo{address}{New York}, \bibinfo{year}{1995}).

\bibitem[{\citenamefont{Butcher}(2008)}]{Butcher:2008jb}
\bibinfo{author}{\bibfnamefont{J.~C.} \bibnamefont{Butcher}},
  \emph{\bibinfo{title}{Numerical Methods for Ordinary Differential Equations}}
  (\bibinfo{publisher}{John Wiley \& Sons}, \bibinfo{address}{Chichester},
  \bibinfo{year}{2008}), \bibinfo{edition}{2nd} ed.

\bibitem[{\citenamefont{Press et~al.}(1992)\citenamefont{Press, Flannery,
  Teukolsky, and Vetterling}}]{Press:1992nr}
\bibinfo{author}{\bibfnamefont{W.~H.} \bibnamefont{Press}},
  \bibinfo{author}{\bibfnamefont{B.~P.} \bibnamefont{Flannery}},
  \bibinfo{author}{\bibfnamefont{S.~A.} \bibnamefont{Teukolsky}},
  \bibnamefont{and} \bibinfo{author}{\bibfnamefont{W.~T.}
  \bibnamefont{Vetterling}}, \emph{\bibinfo{title}{Numerical Recipes: The Art
  of Scientific Computing}} (\bibinfo{publisher}{Cambridge University Press},
  \bibinfo{address}{Cambridge}, \bibinfo{year}{1992}).

\bibitem[{\citenamefont{Kosloff and Tal-Ezer}(1993)}]{Kosloff:1993kt}
\bibinfo{author}{\bibfnamefont{D.}~\bibnamefont{Kosloff}} \bibnamefont{and}
  \bibinfo{author}{\bibfnamefont{H.}~\bibnamefont{Tal-Ezer}},
  \bibinfo{journal}{Journal of Computational Physics}
  \textbf{\bibinfo{volume}{104}}, \bibinfo{pages}{457} (\bibinfo{year}{1993}).

\bibitem[{\citenamefont{Field et~al.}(2010)\citenamefont{Field, Hesthaven, and
  Lau}}]{Field:2010xn}
\bibinfo{author}{\bibfnamefont{S.~E.} \bibnamefont{Field}},
  \bibinfo{author}{\bibfnamefont{J.~S.} \bibnamefont{Hesthaven}},
  \bibnamefont{and} \bibinfo{author}{\bibfnamefont{S.~R.} \bibnamefont{Lau}}
  (\bibinfo{year}{2010}), \eprint{1001.2578}.

\bibitem[{\citenamefont{Abramowitz and Stegun}(1972)}]{Abramowitz:1970as}
\bibinfo{author}{\bibfnamefont{M.}~\bibnamefont{Abramowitz}} \bibnamefont{and}
  \bibinfo{author}{\bibfnamefont{I.~A.} \bibnamefont{Stegun}},
  \emph{\bibinfo{title}{Handbook of Mathematical Functions with Formulas,
  Graphs, and Mathematical Tables}} (\bibinfo{publisher}{Dover},
  \bibinfo{address}{New York}, \bibinfo{year}{1972}).

\end{thebibliography}

\end{document}